\documentclass{caosp306}

\usepackage{graphicx}

\usepackage{natbib}
\usepackage{graphicx}
\usepackage{xcolor}
\usepackage{amssymb}

\newcommand{\mo}{\,$M_{\odot}$}
\newcommand{\lo}{\,$L_{\odot}$}
\articleNo{123}
\pubyear{2019}
\volume{49}
\volnumber{1}
\firstpage{19}
\received{March 22, 2019}
\accepted{April 2, 2019}
\def\BibTeX{{\rm B\kern-.05em{\sc i\kern-.025em b}\kern-.08em
             T\kern-.1667em\lower.7ex\hbox{E}\kern-.125emX}}
\begin{document}
%
\hauthor{M.\,Seker\'{a}\v{s} {\it et al.}}
\title{Photometry of Symbiotic Stars - XIV}
%
\author{
        M.\,Seker\'{a}\v{s}\inst{1}
      \and 
        A.\,Skopal\inst{1}
      \and 
        S.\,Shugarov\inst{1,2}
      \and
	N.\,Shagatova\inst{1}				
      \and
	E.\,Kundra\inst{1}
      \and
	R.\,Kom\v{z}\'{i}k\inst{1}
      \and
	M.\,Vra\v{s}\softt\'{a}k\inst{3}
      \and
        S.\,P.\,Peneva\inst{4}
      \and  
        E.\,Semkov\inst{4}
      \and
	R.\,Stubbings\inst{5}
       }
\institute{
           \lomnica, \email{skopal@ta3.sk}
	 \and 
            P. K. Sternberg Astronomical Institute, M. V. Lomonosov 
            Moscow State University, Russia\\ 
	 \and 
            Private observatory, 03401 Liptovsk\'{a} \v{S}tiavnica, 
            The Slovak Republic\\
         \and
	    Institute of Astronomy and National Astronomical Observatory, 
            Bulgarian Academy of Sciences, 72 Tsarigradsko Shose blvd., 
            BG-1784 Sofia, Bulgaria 
         \and 
            Tetoora Observatory, 2643 Warragul-Korumburra Road, 
            Tetoora Road, Victoria 3821, Australia\\ 					
				}
%
%
\date{March 31, 2019}

\maketitle

\begin{abstract}
We present new multicolour $UBVR_{\rm C}I_{\rm C}$ photometric
observations of symbiotic stars, EG~And, Z~And, BF~Cyg, CH~Cyg, 
CI~Cyg, V1016~Cyg, V1329~Cyg, AG~Dra, RS~Oph, AG~Peg, AX~Per, 
and the newly discovered (August 2018) symbiotic star HBHA~1704-05, 
we carried out during the period from 2011.9 to 2018.75. 
Historical photographic and visual/$V$ data were collected 
for HBHA~1704-05, FG~Ser and AE~Ara, AR~Pav, respectively. 
The main aim of this paper is to present our original observations 
of symbiotic stars and to describe the most interesting features 
of their light curves. For example, periodic variations, rapid 
variability, minima, eclipses, outbursts, apparent changes of 
the orbital period, etc. 
Our measurements were obtained by the classical photoelectric 
photometry (till 2016.1) and the CCD photometry. Main results 
of our monitoring program are summarized and some specific 
characteristics are pointed out for future 
investigation.\footnote{Data are at 
https://www.astro.sk/caosp/Eedition/FullTexts/vol49no1/pp19-66.dat/} 
\end{abstract}
\section{Introduction}
\label{intr}
Symbiotic stars are the widest interacting binary systems that 
comprise a cool giant as the donor and a white dwarf (WD) 
accreting from the giant's wind \citep[][]{1967SvA....11....8B,
1986syst.book.....K,1999A&AS..137..473M}. 
Their orbital periods run from hundreds of days (S-type 
systems containing a normal giant) to a few times 10--100 years 
(D-type systems containing a Mira variable)\footnote{see 
\cite{1975MNRAS.171..171W} for details of this classification}. 
The former case is given by our knowledge of the photometric 
and/or spectroscopic orbital elements 
\citep[e.g.][]{2000A&AS..146..407B}, while the latter is based 
on the long-term spectroscopic observations and/or imaging 
\citep[e.g.][]{2002A&A...391..999P,2006ApJ...637L..49M}. 
However, they are, in general, unknown. 

The accreting WD represents a strong source of the supersoft 
X-ray to UV radiation ($T_{\rm WD} \ga 10^5$\,K, 
$L_{\rm WD} \sim 10^1-10^4$\lo) in the binary 
\citep[][]{1991A&A...248..458M,2005A&A...440..995S,
           2009A&A...507.1531S,2013A&A...556A..85G,
           2016MNRAS.461.3599R,2017A&A...604A..48S}
that ionizes a fraction of the wind from the giant giving rise 
to the nebular emission 
\citep[e.g.][]{1966SvA....10..331B,1984ApJ...284..202S,
               1984ApJ...279..252K,1987A&A...182...51N}. 
This configuration represents the so-called {\em quiescent phase} 
of symbiotic binary, during which the symbiotic system releases 
its energy approximately at a constant rate and temperature. 
Corresponding optical light curves (LC) are characterized with 
a well pronounced periodic wave-like variation along the orbit. 
Analyzes and discussions of this fundamental variability in 
the optical continuum of symbiotic stars were provided by 
\cite{1966SvA....10..331B}, 
\cite{1970Afz.....6...49B}, 
\cite{1986syst.book.....K}, 
\cite{2001A&A...366..157S}, 
\cite{2008JAVSO..36....9S} and 
\cite{2010PASP..122...35J}. 

Sometimes symbiotic stars experience unpredictable outbursts 
characterized by $\sim 1-3$\,mag (multiple) brightening(s) in 
the optical, observed on the timescale of a few months to 
years/decades \citep[see e.g., historical LCs of FN~Sgr, 
Z~And, AX~Per of][]{2005A&A...440..239B,2008MNRAS.385..445L,
2011A&A...536A..27S} with signatures of a mass-outflow 
\citep[e.g.,][]{1995ApJ...442..366F,2000AstL...26..162E,
                2006A&A...457.1003S,2011PASP..123.1062M,
                2013A&A...556A..85G}. 
Outbursts of symbiotic stars are called `Z And-type' outbursts, 
as they were observed in the past for a prototype of the class 
of symbiotic stars -- Z~And \citep[][]{1986syst.book.....K}. 
We name this stage as the {\em active phase}. 

According to the basic composition of symbiotic binaries 
(see above), modelling the spectral energy distribution 
confirmed the presence of three basic components of radiation 
in their spectrum -- two stellar from the giant and the hot 
component and one nebular from the ionized circumbinary 
environment. Their contributions in the optical rival each 
other and are different for different objects, and variable 
due to activity and/or the orbital phase 
\citep[see Figs.~2--22 of][]{2005A&A...440..995S}. 
During active phases, the ionization structure of symbiotic 
binaries changes significantly 
\citep[e.g.][]{2012A&A...548A..21C}, which 
causes dramatic changes of the LC profiles with respect 
to the wave-like variation during the quiescence. In special 
cases, transient narrow minima (eclipses) can emerge in 
the LC for systems with a high orbital inclination 
\citep[e.g.][]{1991BCrAO..83..104B}, whose position and 
profile is a function of the stage of the activity 
\citep[][]{1998A&A...338..599S}. 

With respect to the above mentioned principal behaviour of 
symbiotic stars, their LCs bear a great deal of information 
about the location and physical properties of the radiative 
sources in the system. 
Above all, the multicolour photometry represents an important 
complement to spectroscopic observations and those obtained at 
other wavelengths. Their monitoring also plays an important 
role in discoveries of unpredictable outbursts of symbiotic stars. 

In this paper, we present results of our long-term monitoring
programme of photometric observations of selected symbiotic 
stars, originally launched by \cite{1989IBVS.3364....1H}. 
It continues its part XII \citep[][]{2007AN....328..909S} 
and XIII \citep[][]{2012AN....333..242S} by collecting new data 
obtained during the period from November, 2011 to December 
2018. 
Their acquisition and reduction are introduced in Sect.~2. 
Sect.~3 describes the most interesting features of the LCs 
that deserve further investigation. 
Conclusions are found in Sect.~\ref{s:concl}. 
\begin{table}[p!t]
\begin{center}
\caption{
  Magnitudes of comparison stars used for our targets. The denotation 
  of the stars is adapted from \cite{2006A&A...458..339H}, if not 
  specified otherwise. }
\label{tab:comp}
\begin{tabular}{cccccc}
\hline
\hline
Star &  $V$  & $B$ & $U-B$ & $V-R_{\rm C}$ &$R_{\rm C}-I_{\rm C}$\\
\hline
\multicolumn{6}{c}{Comparison stars in the field of EG\,And} \\
\hline
a      &  8.574 & 10.114 & 1.965 & 0.819 &  0.759  \\
b      & 10.118 & 11.145 & 0.822 & 0.539 &  0.495  \\
\hline
\multicolumn{6}{c}{Comparison stars in the field of Z\,And} \\
\hline
$\beta$  &  9.044 &  9.474  & 0.095 & 0.292 & 0.202 \\
$\gamma$ &  9.229 &  10.549 & 1.229 & 0.744 & 0.681 \\
a        & 12.748 & 13.525  & 0.375 & 0.437 & 0.402 \\
b        & 14.205 & 14.765  & 0.054 & 0.337 & 0.352 \\
c        & 14.083 & 15.200  & 0.874 & 0.624 & 0.583 \\
d        & 14.913 & 15.791  & 0.293 & 0.488 & 0.473 \\
\hline
\multicolumn{6}{c}{Comparison stars in the field of BF\,Cyg} \\
\hline
a        & 11.159 & 11.449  & 0.091 & 0.173 & 0.208 \\
b        & 12.417 & 12.708  & 0.182 & 0.155 & 0.173  \\
\hline
\multicolumn{6}{c}{Comparison stars in the field of CH\,Cyg} \\
\hline
b        &  9.475 & 10.021  &  0.079 & 0.349 & 0.293 \\
d        & 10.227 & 11.288  &  0.810 & 0.564 & 0.453 \\
e        & 10.852 & 12.260  &  1.693 & 0.790 & 0.637 \\
f        & 12.044 & 12.679  &  0.183 & 0.381 & 0.332 \\
\hline
\multicolumn{6}{c}{Comparison stars in the field of CI\,Cyg} \\
\hline
a        &  8.831 &  9.445 & -0.166 & 0.425 & 0.269 \\
b        & 11.722 & 11.996 &  0.198 & 0.159 & 0.173 \\
c        & 12.353 & 13.170 &  0.073 & 0.376 & 0.368 \\
\hline
\multicolumn{6}{c}{Comparison stars in the field of V1016\,Cyg} \\
\hline
a        & 12.314 & 12.866  &  0.085 & 0.334 & 0.315 \\
b        & 12.887 & 13.292  &  0.223 & 0.237 & 0.263 \\
\hline
\multicolumn{6}{c}{Comparison star in the field of V1329\,Cyg} \\
\hline
b        & 12.092 & 13.445  & 1.285  & 0.724 & 0.646  \\ 
\hline
\multicolumn{6}{c}{Comparison stars in the field of AG\,Dra} \\
\hline
a        & 10.459 & 11.018 &   0.015 & 0.333 &      -  \\
b        & 11.124 & 11.857 &   0.183 & 0.416 &  0.330  \\
c        & 11.699 & 12.244 &  -0.042 & 0.335 &  0.294  \\
d        & 11.312 & 12.594 &   1.135 & 0.692 &  0.580  \\
f        & 13.221 & 13.707 &   0.045 & 0.321 &  0.257  \\
\hline
\multicolumn{6}{c}{Comparison stars in the field of Draco~C1$^a$} \\
\hline
 4       & 15.363 & 16.328 & 0.620   & 0.517 & 0.481  \\
 7       & 17.447 & 18.238 & 0.298   & 0.465 & 0.430  \\
 9       & 17.974 & 18.981 & 0.309   & 0.585 & 0.581  \\  
11       & 19.339 & 20.123 & 0.114   & 0.445 & 0.569  \\
12       & 19.787 & 20.587 &  --     & 0.407 & 0.643  \\
\hline   
\end{tabular}
\end{center} 
\end{table}  
\addtocounter{table}{-1}
\begin{table}[!ht]
\small   
\begin{center}
\caption{continued}
\begin{tabular}{lccccc}
\hline
Star &  $V$  & $B$ & $U$-$B$ & $V$-$R_{\rm C}$ &$V$-$I_{\rm C}$\\
\hline
\multicolumn{6}{c}{Comparison stars in the field of RS\,Oph} \\
\hline
$\alpha$ &  9.307 & 10.543 &  0.880 & 0.706 &  0.650  \\
$\beta$  & 11.494 & 12.198 &  0.160 & 0.450 &  0.379  \\
a        & 13.255 & 14.272 &  0.377 & 0.660 &  0.652  \\
\hline
\multicolumn{6}{c}{Comparison stars in the field of AG\,Peg} \\
\hline
a        & 10.428 & 11.059 &   0.178 & 0.374 &  0.308 \\
b        & 10.672 & 11.505 &   0.528 & 0.508 &  0.430 \\
c        & 11.204 & 11.708 &  -0.030 & 0.330 &  0.302 \\
d        & 11.650 & 12.628 &   0.715 & 0.537 &  0.503 \\
\hline
\multicolumn{6}{c}{Comparison stars in the field of AX\,Per} \\
\hline
$\alpha$ & 10.201 & 10.290 & -0.177 & 0.023 &  0.043  \\
$\beta$  & 11.156 & 12.323 &  0.993 & 0.608 &  0.531  \\
a        & 12.212 & 12.854 &  0.166 & 0.384 &  0.366  \\
b        & 13.012 & 13.025 & -0.448 & 0.004 &  0.057  \\
c        & 12.124 & 13.141 &  0.700 & 0.544 &  0.515  \\
\hline
\multicolumn{6}{c}{Comparison stars in the field of HBHA\,1704-05$^b$} \\
\hline
120 & 12.627(30)& 12.525(10)& 11.936(5)& 11.581(10)& 11.239(10)\\
123 & 13.020(30)& 12.873(10)& 12.260(5)& 11.903(10)& 11.549(10)\\
 C  & 11.981(40)& 11.357(10)& 10.519(5)& 10.039(10)&  9.626(10)\\
\hline
\hline
\end{tabular}
\end{center}
{\bf Notes:}
$^a$ according to \cite{2000A&AS..143..343H}; \\
$^b$ C = TYC~1620-2479 
($\alpha_{2000} = 19^{\rm h}54^{\rm m}55^{\rm s},~
 \delta_{2000} = +17^{\circ}23^{\prime}36^{\prime\prime}$, 
 Sect.~\ref{s:hbha}). 
\end{table}
\section{Observations and data reduction}
\label{s:obs}
Our photometric measurements were carried out at different 
observatories using different devices and techniques. 
Multicolour photometric measurements were obtained mostly at 
the Star\'{a} Lesn\'{a} observatory (pavilions G1 and G2) 
operated by the Astronomical Institute of the Slovak Academy 
of Sciences using two 60 cm, f/12.5 Cassegrain telescopes and 
one 18 cm, f/10 Maksutov-Cassegrain telescope. 
New observations cover the period between November 2011 and 
December 2018. 

(i) 
In the pavilion G2, classical photoelectric $UBV$ observations 
in the standard Johnson system were obtained by a single-channel 
photoelectric photometer mounted in the Cassegrain focus of 
the 60 cm reflector 
\citep[see][in detail]{2004CoSka..34...45S,2014CoSka..44...77V,
                       2014CoSka..44...91V,2015CoSka..45...99V}. 
We performed these measurements till February 2016, when they were 
completely replaced by the observations with CCD detectors. 
Magnitudes obtained with photoelectric photometer were 
determined using the comparison stars listed in Table~1 of 
\cite{2012AN....333..242S}. 
Maximum internal uncertainty of these night-means (rms) is of 
a few times 0.01 in the $B,V$ filters and up to $\sim 0.05$\,mag 
in the $U$ passband. 

(ii) 
CCD photometric measurements were obtained in the standard 
Johnson-Cousins $UBVR_{\rm C}I_{\rm C}$ system, using three 
different CCD cameras. Until January 2016, one of the 60 cm 
telescope was equipped with Moravian Instruments G4-9000 CCD camera 
($3056\times 3056$\,px, pixel size: $12\mu\rm m\times12\mu\rm m$), 
for $BVR_{\rm C}I_{\rm C}$ photometry. 
Currently, we are using the FLI~ML3041 CCD camera 
($2048\times 2048$\,px, pixel size: $15\mu\rm m\times15\mu\rm m$, 
scale: 0.4\,arcsec/px, FoV: $14^{\prime} \times 14^{\prime}$) 
mounted at the 60 cm telescope in the G2 pavilion to measure 
$UBVR_{\rm C}I_{\rm C}$ magnitudes, and the SBIG ST10 MXE CCD camera 
($2184\times1472$\,px, pixel size: $6.8\mu\rm m\times6.8\mu\rm m$) 
in the Cassegrain focus of the 18 cm telescope in the G1 pavilion 
for $BVR_{\rm C}I_{\rm C}$ photometric observations. 
As the 18 cm telescope is fixed with the 60 cm one for the 
mid-resolution ($R\sim 11000$) echelle spectroscopy, it is 
also used to obtain simultaneous photometry for absolute flux 
calibration of the spectroscopic observations. 

Standard reduction of the CCD images was done using the 
{\small IRAF}-package software\footnote{IRAF is written and 
supported by the National Optical Astronomy Observatories 
(NOAO) in Tucson, Arizona}. The corresponding magnitudes 
were obtained with the aid of the comparison stars of 
\cite{2006A&A...458..339H}, listed in Table~\ref{tab:comp}. 
Each value represents the average of 
individual measurements during a night. For the conversion of 
the observed magnitudes to the standard $UBVR_{\rm C}I_{\rm C}$ 
system, we used transformation coefficients determined from 
photometric measurements of M67 star cluster. We re-measure 
the system once a year, but also after each change of 
instrumentation including coating the mirrors. 

(iii) 
From 2016.9, some multicolour $BVR_{\rm C}I_{\rm C}$ photometry 
was performed at the private observatory in Liptovsk\'a 
\v{S}tiavnica by one of us (M.V.) using the 35 cm, f/4.5 
Newtonian reflector equipped with Moravian Instruments 
G2-1600 CCD camera (CCD chip: KAF 1603ME, 
$1536\times1024$\,px, pixel size: $9 \times9\,\mu\rm m$). 
The CCD images were reduced with the {\small MuniWin}-package 
software\footnote{http://c-munipack.sourceforge.net/}. 
Resulting magnitudes agree with the values from G1/G2 stations 
(see point (ii)) within the errors, which are in the range of 
$0.02-0.03$\,mag in the $BVR_{\rm C}I_{\rm C}$ filters. 
Because the camera has only a weak and unevenly sensitivity in 
the $U$ passband, we present $U$ magnitudes only of those 
symbiotic stars, for which agreement with our data is within 
$\pm 0.05$\,mag. 

(iv)
Additional $UBVR_{\rm C}I_{\rm C}$ observations 
were performed using the 50/70/172 cm Schmidt telescope of 
the National Astronomical Observatory Rozhen, Bulgaria. 
A CCD camera FLI PL16803 with the chip size $4096 \times 4096$\,px, 
$73.80 \times 73.80$ arcmin field and the scale 
1.08 arcsec/px was used. 
%
Some observations were also carried out with the VersArray 1300\,B 
CCD camera (1340$\times$1300\,px, pixel size: 
20\,$\mu$m $\times $20\,$\mu$m, scale: 0.258 arcsec/px) on 
the 2 m and 1.3 m telescopes. 

Maximum internal uncertainties of our CCD measurements were 
in most cases comparable to those determined for the photoelectric 
$UBV$ photometry (see the point (i) above). In the $R_{\rm C}$ 
and $I_{\rm C}$ passband the night-means (rms) errors were 
$\lesssim$0.05\,mag. However, in most cases the errors are 
given in tables (see on-line data). 

(v)
The photographic magnitudes of FG~Ser and HBHA~1704-05 
were determined using the digitized plates from the archive of 
the Sternberg Astronomical Institute of the Moscow University 
(SAI). The plates were obtained with the 40 cm astrograph at 
the SAI Crimean Observatory. 
To extract the LC of these objects, we reprocessed the original 
scans using the latest version of the {\small VaST}-package 
software\footnote {http://scan.sai.msu.ru/vast/} featuring 
the improved photographic aperture photometry calibration 
technique 
\citep[][]{2005MNRAS.362..542B,2016arXiv160503571S}. 

(vi) 
In addition, visual magnitude estimates of symbiotic stars 
AE~Ara and AR~Pav were performed at the Tetoora Observatory in 
Australia (R.S.) with the 55 cm, f/3.8 Dobsonian telescope, from 
2010. Between 2007 and 2010, estimates of these stars were carried 
out by Albert Jones\footnote{1920--2013} at the Carter Observatory, 
Wellington 1, New Zealand, using his private 32 cm, f/5 reflector. 
Comparing visual estimates with independently obtained $V$ 
magnitudes (e.g., from ASASSN survey), we determined their 
uncertainties to $\sim0.3$\,mag for AE~Ara (see Figs.~\ref{aefit} 
and \ref{aeara} here, and Fig.~3 of \cite{2007AN....328..909S}). 
For AR~Pav we found uncertainties to be even smaller, being 
of $\sim0.2$\,mag (Fig.~\ref{arpav} here and Figs.~1 and 10 in 
\cite{2001IBVS.5195....1S} and \cite{2007AN....328..909S}). 
%

Tables with data for individual objects are only available 
on-line at\\
https://www.astro.sk/caosp/Eedition/FullTexts/vol49no1/pp19-66.dat/ . 

\section{Light curves of the measured objects}
\label{s:lcs}
\subsection{EG~And}
\label{ss:egand}

\begin{figure}[p!t]
%
\begin{center}
\resizebox{\hsize}{!}{\includegraphics[angle=-90]{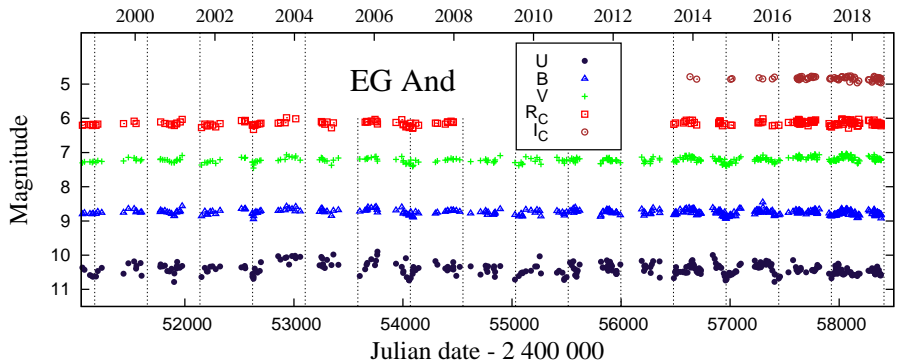}}
\end{center}
\caption{
$UBVR_{\rm{C}}I_{\rm{C}}$ LCs of EG~And. Vertical lines 
represent times of the inferior spectroscopic conjunction of 
the giant according to the ephemeris of 
\cite{2016AJ....152....1K}, 
$JD_{\rm sp. conj.} = 2\,450\,208.108(\pm 0.672) + (482.5\pm 1.3)\times E$. 
New observations are displayed with those of 
\cite{2012AN....333..242S}. 
}
\label{egand}
\end{figure}

EG~And is one of the brightest symbiotic stars. To date, no 
outburst was detected during its observational history 
\citep[][and Fig.~\ref{egand} here]{1997A&A...318...53S,
2010PASP..122...35J}. 
It is an eclipsing binary consisting of M2.4\,{\small III} giant 
and a hot star with a relatively low temperature of 
$70\,000 - 95\,000\,\rm K$ in comparison with other symbiotic 
stars \citep[][]{1985ApJ...295..620O,1987AJ.....93..938K,
                 1991A&A...248..458M,2005A&A...440..995S}. 
The orbital period was most recently determined by 
\cite{2016AJ....152....1K} to $P_{\rm orb}=483.3\pm1.6\rm$ days 
from radial velocities of the cool component derived from echelle 
spectra observed during more than 20 years. Including their data 
to all the previously published, they attained an average orbital 
period of $P_{\rm orb}=482.5\pm1.3\rm$ days. 
%

\begin{figure}[p!t]
\centerline{
\includegraphics[angle=-90, width=0.9\textwidth,clip=]{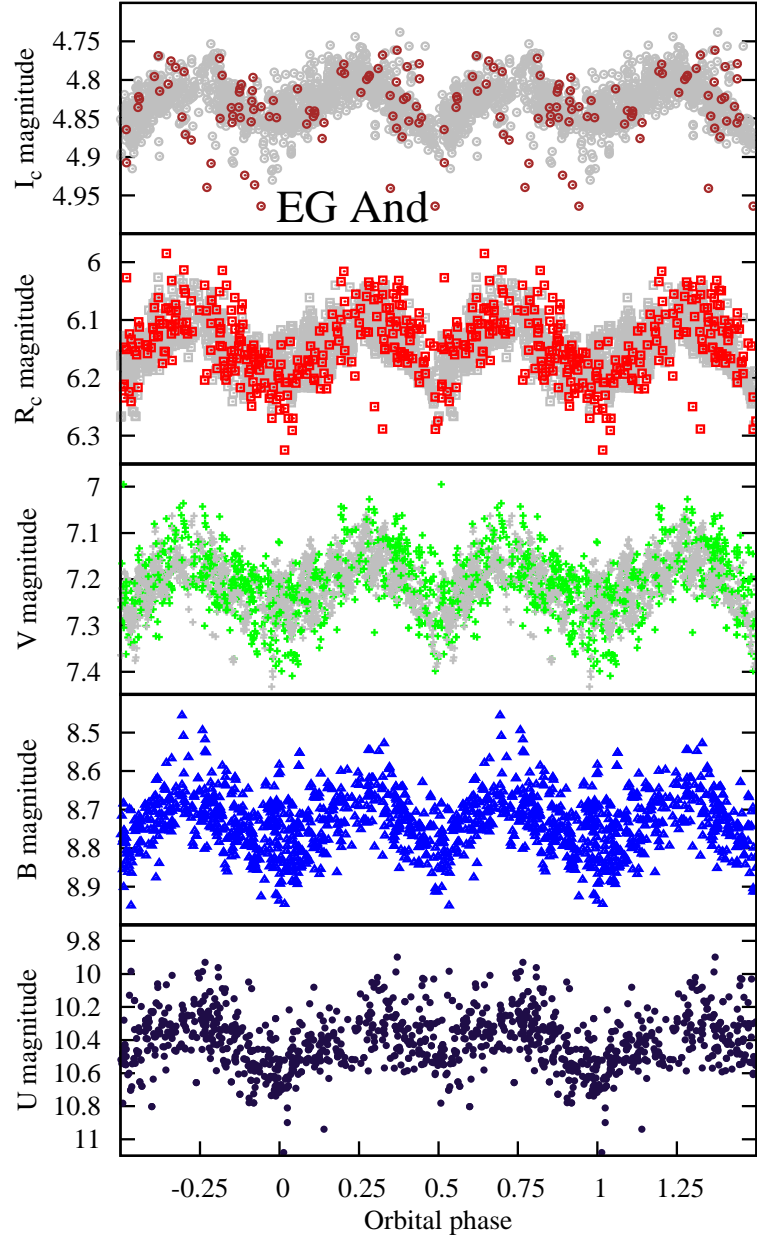}}
\caption{
Phase diagrams of our $UBVR_{\rm{C}}I_{\rm{C}}$ magnitudes for EG~And. 
Compared are $VRI$ data of \cite{2001PASP..113..983P} shifted by 
an appropriate constant (in grey). The large scatter of the data 
is caused by several types of variabilities (see text). 
}
\label{egand_dw}
\end{figure}

Although EG~And persists in a quiescent phase, we can identify 
a few types of light variability in it's LC. 
Figure~\ref{egand_dw} shows a double-wave light variation along 
the orbital motion with minima around times of the spectroscopic 
conjunctions of the binary components. This variability dominates 
the light curve of EG~And in all passbands. 
The presence of the secondary minimum around the orbital phase 0.5 
in the $I_{\rm C}$ filter is confirmed by the long-term 
photometry in the Johnson's $I$ filter as published by 
\cite{2001PASP..113..983P}. 

Based on modelling the $B$ and $V$ LCs of EG~And, 
\cite{1997MNRAS.291...54W} attributed the double-wave LC 
to the tidal distortion of the giant.  
On the other hand, \cite{2001A&A...366..157S} proposed a model, 
where a partially optically thick, non-symmetrical symbiotic 
nebula can also generate the observed orbitally-related 
double-wave light variation by producing different contributions 
of its emission into the line of sight at different orbital 
phases. This interpretation does not claim the Roche-filling 
giant, recently confirmed by \cite{2016AJ....152....1K}. 
Finally, we note that \cite{2010PASP..122...35J} didn't find 
any ellipsoidal variability in their photographic data obtained 
during 35 years of observations. However, the accuracy of their 
estimates of $0.1$\,mag and the presence of additional 
$\sim0.1$\,mag variation in $B$ on the time scale of tens 
of days (see below) could affect the visibility of the 
secondary minimum. 
\begin{figure}[p!t]
\centerline{
\includegraphics[angle=-90, width=1.0\textwidth,clip=]{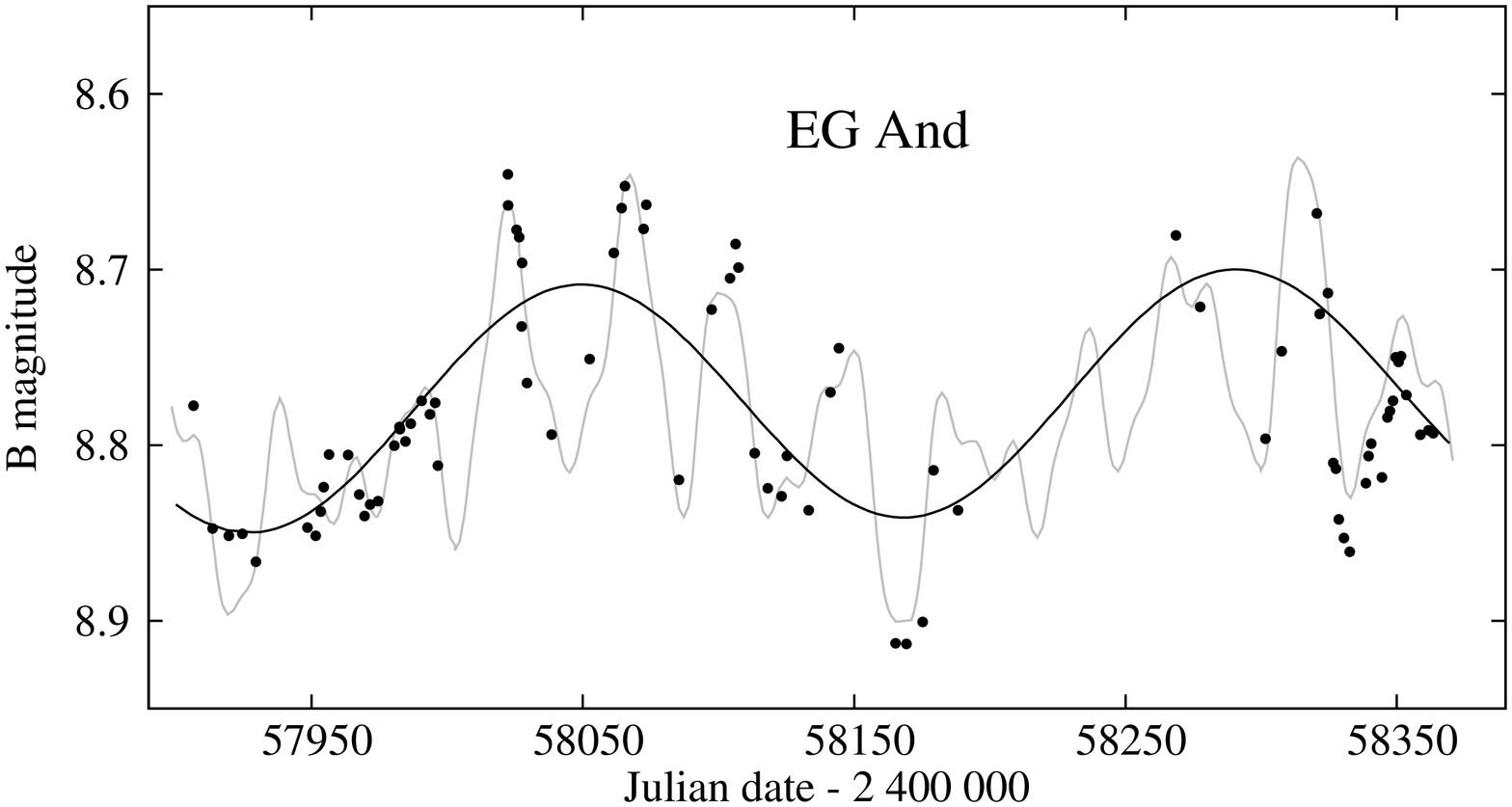}}
\caption{
A part of the $B$ LC of EG~And, where a light variation with peaks 
separated by several tens of days is clearly recognized. 
The grey line represents a model including periods: 241 (half of 
the $P_{\rm orb}$ -- black line), 39, 46, and 28 days (see text).
}
\label{egand_puls}
\end{figure}

The large scatter of data in all filters shown in the phase 
diagram (Fig.~\ref{egand_dw}) represents real light 
variations \citep[][]{1997A&A...318...53S,2012AN....333..242S}. 
In our data, this variability is best evident in $B$ and $V$ 
after May 2017 ($\sim JD\sim2\,457\,900$, Fig.\,\ref{egand_puls}). 
The peaks are separated by $\sim40$ days. Using Fourier analysis 
on the time interval between May 2017 and March 2018 we 
identified two periods of $\sim39$ and $\sim46$ days. 
Also, the wavelet analysis revealed a period with the highest 
WWZ statistic of $\sim42$ days. \cite{2001PASP..113..983P} 
identified a period of 29 days by analyzing the long-term 
$VRI$ photometry of EG~And. \cite{2006PhDT.........4C} 
suggested, that the period of $\sim28$ days with a variation 
of $\sim0.1$\,mag is consistent with small radial 
pulsations of the red giant photosphere. Although we 
examined only the short time interval, we found a $\sim$28 
days period in the $B$ data. When using all our data in $B$ 
together with those of \cite{1997A&A...318...53S} and 
\cite{2012AN....333..242S}, 
we identified a period of $\sim29$ days. It has, however, 
lower significance than that of 42 days, which we did not 
find using all the data. 
The brightness of the star in $U$ filter was observed less 
frequently and these "pulsations" are not so obvious. 
In our data, we found only a period of $\sim27$ days 
after $JD\sim2\,457\,900$. The light curve with $U$ magnitudes 
shows more chaotic behaviour, with variation up to 
$\sim0.5$\,mag. The observed brightenings could be related 
to the enhanced mass loss rate from the giant due to pulsations 
of the giant. 
The enhanced mass loss from the giant increases the 
density of the circumstellar matter and hence the 
accretion rate onto the WD. The enhanced 
accretion can increase the luminosity of the WD, 
and thus also the nebular radiation, which dominates 
the $U$ passband 
\citep[][]{2006PhDT.........4C,2005A&A...440..995S,
           2012AN....333..242S}. 

Finally, there is probably present a very long-period light 
variation on the time-scale of several years. It is apparent 
mainly in the $U$ brightness, which peaks between the years 
1986--1989 \citep[][]{1997A&A...318...53S} and 2004-2006 
\citep[][]{2012AN....333..242S}. This light variation has the 
highest amplitude in $U$ ($\pm 0.1$\,mag) and lowest in 
$R_{\rm C}$ and $I_{\rm C}$ band ($\pm0.025$\,mag). 
%
%
\begin{figure}[p!t]
\centerline{
\includegraphics[angle=-90, width=1.0\textwidth,clip=]{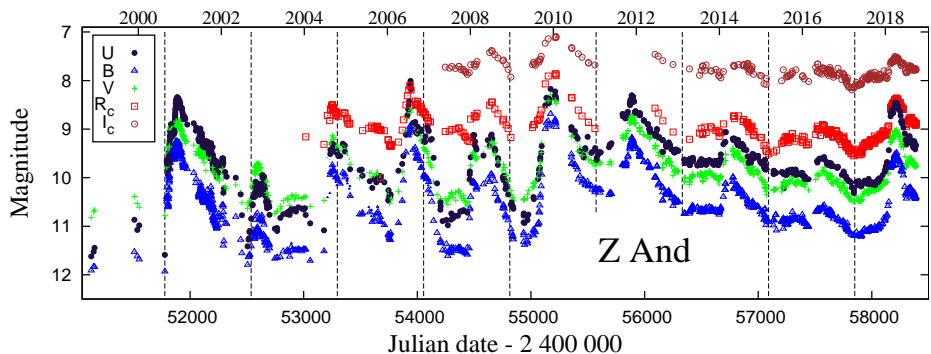}}
\caption{
As in Fig.~\ref{egand}, but for Z~And. Times of the inferior 
conjunctions of the giant (vertical lines) are according to 
the ephemeris of \cite{2000AJ....120.3255F}, 
$JD_{\rm sp. conj.} = 2\,450\,260.2(\pm 5.4) + (759.0\pm 1.9)\times E$.
Data from \cite{2007AN....328..909S,2012AN....333..242S} 
are supplemented by our data from 2011.9. 
}
\label{zand}
\end{figure}
\subsection{Z~And}

Z~And is considered to be a prototype of the class of 
symbiotic stars. 
The binary consists of a late-type M4.5 III red giant and 
a WD accreting from the giant’s wind on the 758-day orbit 
\citep[e.g.][]{1989A&A...213..137N,2000AJ....120.3255F}. 
The current active phase has begun in September 2000 
\citep[][]{2000IBVS.5005....1S}, being followed with a series 
of outbursts with the main optical maxima in December 2000, 
July 2006, December 2009 and December 2011 (see Fig.~\ref{zand}). 
At/after the 2006 and 2009 maxima, highly collimated bipolar 
jets were detected as the satellite emission components to 
H$\alpha$ and H$\beta$ lines 
\citep[][]{2006ATel..882....1S,2018ApJ...858..120S}. 

Our new photometry is shown in Fig.~\ref{zand}. 
After the major 2009 outburst, we detected four new main peaks 
in the LC. The first occurred in November 2011 
($JD_{\rm Max}\sim2\,455\,887$) with maximum $U\sim8.3$\,mag). 
Then the brightness was continuously fading until March 2013 
($JD_{\rm Min}\sim2\,456\,353$) to $U\sim9.7$\,mag. 
The next maximum occurred between February and May 2014 at 
$U\sim9.1$\,mag. 
In June 2016, the star's brightness slightly increased to 
$\sim9.6$\,mag, and then declined to the lowest level at 
$U\sim10.3$\,mag, in April 2017. 
In October 2017, Z~And started the recent most pronounced outburst 
since that in 2011. The brightness steeply increased to the maximum 
of $U\sim8.4$\,mag in April 2018, following a decrease to the 
present (December 2018). 
\begin{figure}[p!t]
\centerline{
\includegraphics[angle=-90, width=1.0\textwidth,clip=]{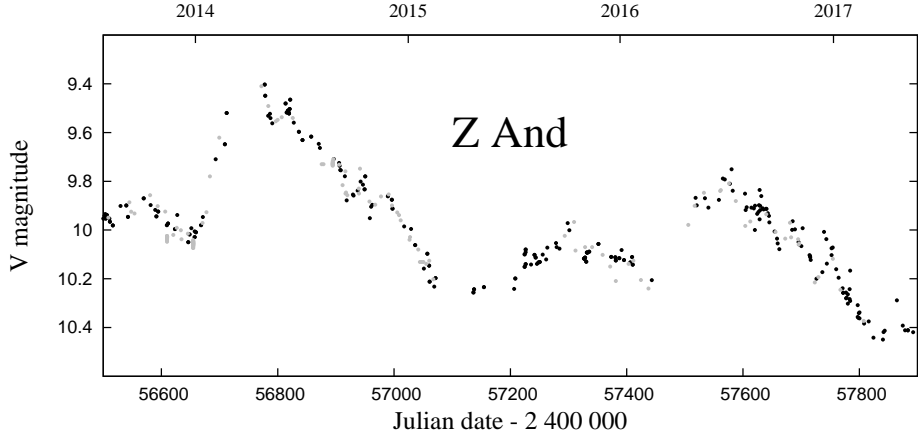}}
\caption{
The $V$ LC of Z~And from the \textsl{AAVSO} database (grey points) 
and our measurements (black points). Observations suggest 
the presence of a 58-d periodic variability. 
}
\label{zand_close}
\end{figure}

The profiles of the observed outbursts since 2011 are very similar. 
The steep increase of the brightness, followed by a slower decline 
and separated from the next increase with a standstill in the 
brightness. 
In April 2017, we observed a short decline by $\sim0.25$\,mag 
in $U$ just before the phase of relatively constant brightness, 
close to the orbital phase $\varphi\sim0$, as observed in the LC 
of BF~Cyg (see Fig.~\ref{outbursts}). 
The $R_{\rm C}$ and $I_{\rm C}$ LCs show a double-wave light 
variation. 

Our data supplement well those from the \textsl{AAVSO} database, 
whereby we can detect a light variation with the period of 
$\sim58$ days, best visible in the $V$ LC (Fig.\,\ref{zand_close}). 
\subsection{AE~Ara}
\label{ss:aeara}

AE~Ara was specified as a symbiotic star on the basis of its 
infrared colours by \cite{1974MNRAS.167..337A}. It is an S-type 
symbiotic star with the orbital inclination of $51^{\circ}$ 
\citep[][]{2010AJ....139.1315F}, thus no eclipses are observed 
in the LC. 
The orbital period of $812 \pm 2$ days was determined by 
\cite{2003ASPC..303..147M} using the visual photometry 
during the 1980s and 1990s decades. 
Using radial velocity measurements from the near-IR spectrum 
obtained during 2001 -- 2009, \cite{2010AJ....139.1315F} found 
the value of $803\pm 9$ days. 

\begin{figure}[p!t]
\centerline{
\includegraphics[angle=-90, width=1.0\textwidth,clip=]{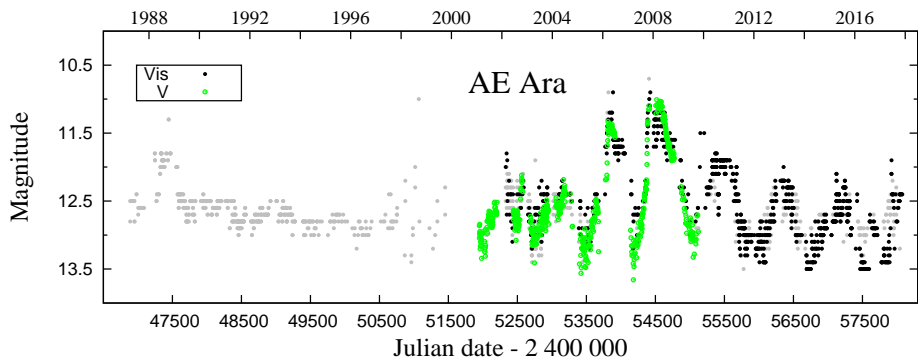}}
\caption{
The LC of AE~Ara consisting of visual estimates from the 
\textsl{AAVSO} database (grey dots), our data (black) and 
$V$ magnitudes from \textsl{ASAS} database (green).}
\label{aeara}
\end{figure}

The published photometric data consist mainly of visual estimates. 
They are in good agreement with photometric $V$ magnitudes 
published by \cite{2007AN....328..909S} and $V$ magnitudes released 
as a part of the ASAS\footnote{All Sky Automated Survey} program 
\citep[][]{2002AcA....52..397P}. 
According to the LC evolution, AE~Ara was in quiescent phase 
until the beginning of 2006, when the system entered a new active 
phase \citep[see Fig.~3 of][ and Fig.~\ref{aeara} here]
{2007AN....328..909S}. 

\begin{figure}
\centerline{
\includegraphics[angle=-90, width=1.0\textwidth,clip=]{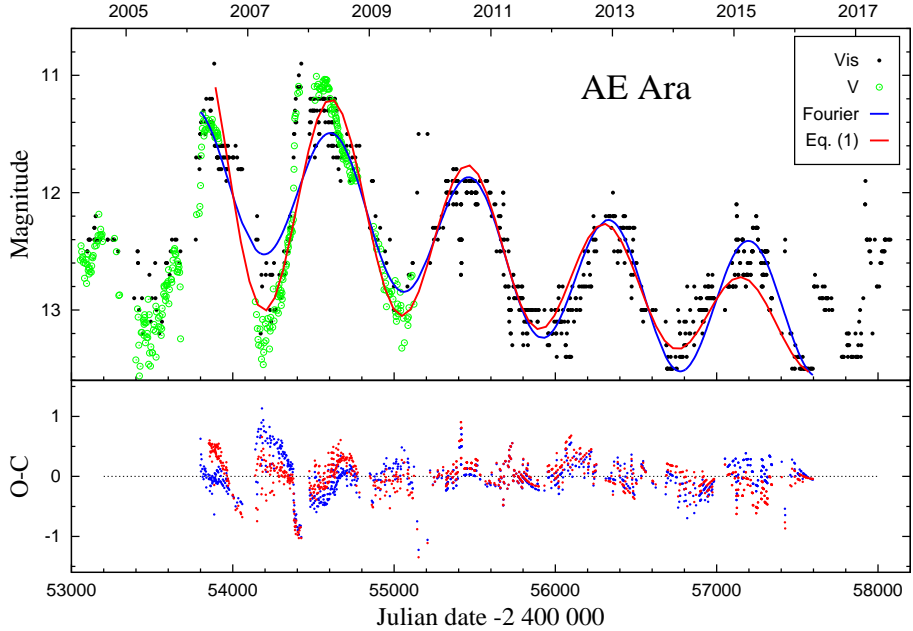}}
\caption{
Top: Our fits to the $V$/visual LC of AE~Ara during its 
transition from the active to quiescent phase (2006.2 -- 2016.6). 
The corresponding orbital period is larger than that determined 
from the quiescent LC (see text). 
Bottom: $O-C$ residuals between the fit from Fourier analysis 
(blue points), Eq.~(\ref{new}) (red points) and the data. 
}
\label{aefit}
\end{figure}

Here we present new visual observations from 2007 
(see Sect.~\ref{s:obs}, point (vi)). 
Figure~\ref{aeara} shows that AE~Ara was in an active phase 
from 2006 to $\sim$2014 with the maximum $V\sim11$\,mag 
at the beginning of 2008 that was followed by a gradual decrease 
throughout a few orbital cycles. 
Applying the Fourier analysis for the \textsl{AAVSO}\footnote{
American Association of Variable Star Observers} data, $V$ 
magnitudes from \textsl{ASAS} and our measurements between 
March 2006 and July 2016, we estimated the orbital period 
of AE~Ara to $\sim860$ days (Fig.~\ref{aefit}). 
%
To confirm this result, we also used a different approach. 
We fitted the data with a function 
\begin{equation}
  m(t)=A\times e^{-\gamma t^2}\sin[(t-t_0)2\pi/P]+v_1t+v_2,
\label{new}
\end{equation}
where the variation of the magnitude $m(t)$ with time $t$ 
in Julian days is described by a sinus function with the phase 
shift $t_0$ and the period $P$. In our approach, the decrease 
of the amplitude $A$ is given by a damping parameter $\gamma$ 
and the star's brightness declines linearly with time, given 
by parameters $v_1$ and $v_2$. The resulting fit (Fig.~\ref{aefit}) 
yields the period $P=850\pm 20$ days and the $60\%$ decline 
in the amplitude variations from 2006.2 to 2016.6. During this time, 
the brightness of the system decreased with a factor of $\sim1.4$. 
Our analysis suggests that the orbital period indicated during the 
transition from the active to the quiescent phase (2006.2 -- 2016.6) 
is larger than that determined during the quiescent phase by 
\cite{2003ASPC..303..147M} and that from radial velocities 
of \cite{2010AJ....139.1315F}. 
%
%
%

Analysis of the minima positions in the historical LCs of 
symbiotic stars revealed apparent changes in their orbital 
periods that are systematic, depending on the level of activity 
of symbiotic binary \citep[][]{1998A&A...338..599S}. 
For example, after the main 1895 outburst of BF~Cyg the minima 
were separated with $\sim$770 days along the decrease of the 
star's brightness, which indicates a longer orbital period than 
given by the spectroscopic orbit 
\citep[757.2 days; see][]{2001AJ....121.2219F}. 
The nature of the larger apparent orbital period determined for 
AE~Ara during the brightness decline after the 2006 outburst 
will be probably the same as for BF~Cyg. 
Understanding of this very interesting effect requires a more 
detailed analysis, which is out of the scope of this paper. 
\subsection{BF~Cyg}

Observational history of the eclipsing symbiotic star BF~Cyg began 
more than 120 years ago 
\citep[][and references therein]{1941BHarO.915...17J}. 
The main feature of the LC is a very slow decrease in the brightness, 
from the $\sim$1894 nova-like outburst to $\sim$1985, when 
the brightness in $B$ decreased by more than 3\,mag 
\citep[see Fig.~1 of][]{1997MNRAS.292..703S}. 
Simultaneously, periodic or eruptive variations in the star's 
brightness were observed. The historical LC was analyzed by 
\cite{2006MNRAS.366..675L}. They identified several periods in 
the optical LC. Beside the orbital period $P_{\rm orb}=757.3$\,days, 
\citep[][]{1970VilOB..27...24P,2001AJ....121.2219F}, they indicated 
a 798.8-d period, which they attributed to the rotation of 
the giant. In addition, they showed that the most pronounced 
outbursts repeat every $\sim6376$ days, predicting a new outburst 
for 2007. 

A new active phase began around the mid of 2006, reaching the peak 
in September 2006 
\citep[][]{2006CBET..596....1M,2006CBET..633....1I,2007AN....328..909S,
           2011PASP..123.1062M}. 
\begin{figure}[p!t]
\centerline{
\includegraphics[angle=-90, width=1.0\textwidth,clip=]{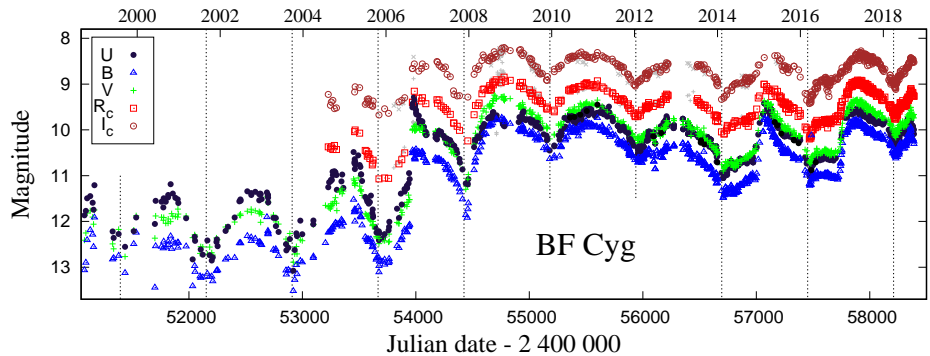}}
\caption{
As in Fig.~\ref{egand}, but for BF~Cyg. Vertical lines represent 
times for conjunctions with the giant in front according to 
the ephemeris of \cite{2001AJ....121.2219F}, 
$JD_{\rm sp. conj.} = 2\,451\,395.2(\pm5.6) + (757.2\pm3.9)\times E$. 
Our new observations are displayed with the data published 
by \cite{2007AN....328..909S,2012AN....333..242S}, 
\cite{2005A&AT...24..447Y} and, for a comparison, data from the 
\textsl{AAVSO} database (grey points). 
}
\label{bfcyg}
\end{figure}

Figure~\ref{bfcyg} shows the multicolour LCs of BF~Cyg from 1999 
to our last observations in December 2018. The light variation 
after the 2006 outburst is modulated by the orbital motion, 
showing broad minima around the inferior conjunction of the giant 
(orbital phase $\varphi = 0$), similar to those observed during 
quiescent phases. 
After the rapid descent to the eclipse in December 2013, the LC 
changed in profile. In 2014 and 2016 the minima were followed 
by a phase of a relatively constant brightness until 
$\varphi\sim0.4$. It is of interest to note, that similar 
light variation was observed also in the LC of AX~Per and 
Z~And (see Fig.~\ref{outbursts}). 
During the 2014 eclipse, BF~Cyg reached the lowest brightness 
since the deep eclipse in 2007. 
Thereafter, the overall brightness of the system was gradually 
increasing. The minima in 2014 ($U\sim11.0$), in 2016 
($U\sim11.0$) and in 2018 ($U\sim10.4$\,mag) occurred 
$\sim$14, $\sim$17 and $\sim$20 days after the predicted inferior 
conjunction of the giant, according to the ephemeris of 
\cite{2001AJ....121.2219F}. 
After that we observed a steep increase of the brightness 
\citep[bursts, see][for the 2016/2017 one]{2017ATel10086....1S}, 
which peaked around $\varphi\sim0.5$. 
A similar profile of the light variation is present in all 
$UBVR_{\rm C}I_{\rm C}$ filters, although in the $I_{\rm C}$ 
filter it is less evident. 
After the most recent minimum in 2018 ($JD_{\rm Min}\sim2\,458\,230$) 
the brightness of BF~Cyg steeply raised to 
$U\sim9.7$\,mag at $JD\sim2\,458\,360$, but consequently dropped 
to $U\sim10$\,mag within $\sim$30 days as given by our last 
measurement. 
\subsection{CH~Cyg}

The photometric variability of CH~Cyg is probably the most 
intriguing from all well-observed symbiotics. The light 
variations are observed on very different time-scales, from 
minutes to decades. 
It is believed that the rapid light variation (flickering, flares 
and bursts) and other types of stochastic variability are connected 
with accretion phenomena in the system. Also possible pulsation 
of both the cool giant and the hot component, orbital motion(s) in 
the system of unknown composition (binary or triple star system?) 
and dust obscuration events make the resulting light and its 
changes even more complex 
\citep[e.g.][]{
1988A&A...198..150M,
1996A&A...308L...9S,
1998MNRAS.297...77I,
2000JAVSO..28..106M,
2001ApJ...560..912C,
2002MNRAS.335.1109S,
2003ApJ...584.1021S,
2006ARep...50...43B,
2006ApJ...647..464B,
2007AstL...33..531T,
2008ARep...52..403B,
2009ApJ...692.1360H,
2009MNRAS.397..325P,
2010ATel.2394....1S,
2010ApJ...710L.132K,
2010ATel.2707....1S,
2012BaltA..21..184S,
2012ATel.4316....1S,
2015EAS....71..107S,
2017ATel10142....1I,
2017Ap.....60..153K,
2017ASPC..510..170M,
2018BlgAJ..28...42S}. 
\begin{figure}[p!t]
\centerline{
\includegraphics[angle=-90, width=1.0\textwidth,clip=]{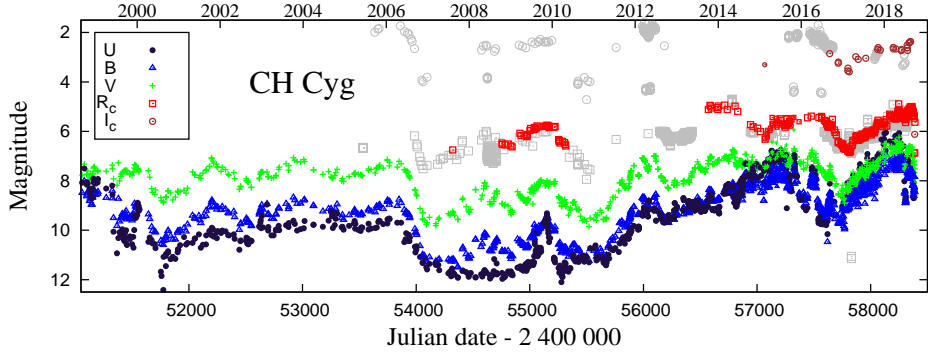}}
\caption{
As in Fig.~\ref{egand}, but for CH~Cyg. New data (from 2011.9) are 
supplemented with those published by 
\cite{2007AN....328..909S,2012AN....333..242S} and the \textsl{AAVSO} 
$R_{\rm C}$ magnitudes (grey points).} 
\label{chcyg}
\end{figure}
\begin{figure}
\centerline{
\includegraphics[angle=-90, width=1.0\textwidth,clip=]{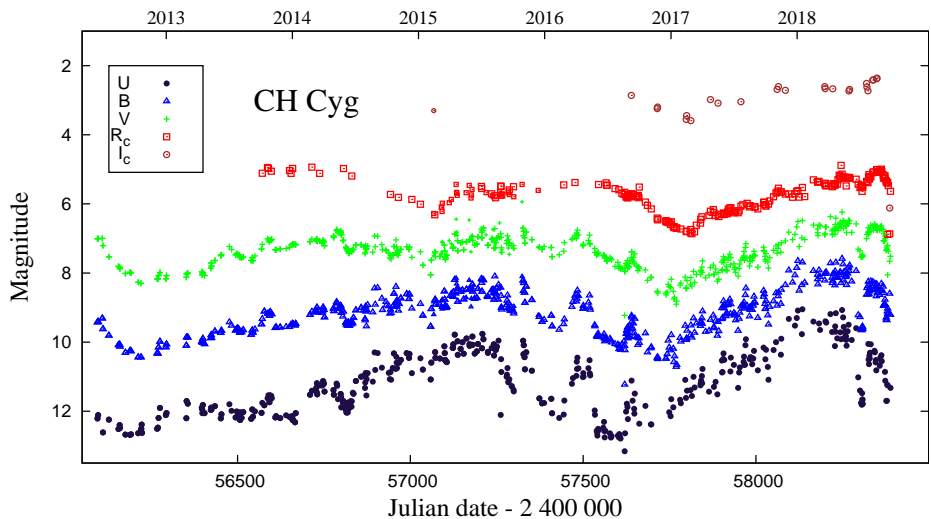}}
\caption{
Detail of Fig.~\ref{chcyg} from 2012.4. The $U$ and $B$ LCs were 
shifted by 3 and 0.7\,mag, respectively, for better visualization.  
}
\label{ch_close}
\end{figure}

Our observations are shown in Figs.~\ref{chcyg} and \ref{ch_close}. 
After the June-December period in 2006, when the LC showed 
a $2$\,mag decline of brightness in all colours 
\citep[][]{2007AN....328..909S}, CH~Cyg began to gradually 
increase its brightness. In 2009, a burst was detected 
simultaneously in the X-rays and the optical $U,B$ passbands 
\citep[][]{2009ATel.2245....1M,2010ATel.2394....1S}. 
Similar bursts were later indicated during October 2015, 
between February and May 2016 and in September 2016. All 
bursts were comparable in amplitude ($\Delta U\sim1.7$\,mag; 
Fig.~\ref{ch_close}). 

From $\sim$2015 we measured a $\gtrsim$3 years lasting wave with 
maxima around July-August 2015 ($U\sim 7$\,mag) and 
March-April 2018 ($U\sim 6.4$\,mag) and the broad minimum at 
the end of 2016 with $U\sim9.7$\,mag. Times of minima depend on 
the colour. 
The largest difference of $\sim190$ days is between the minimum 
in $U$ ($JD\sim2\,457\,620$) and $R_{\rm C}$ ($JD\,2\,457\,814$) 
(Fig.~\ref{ch_close}), although the $U$ LC is rather complex 
during the second half of 2016. It is of interest to note that 
the 2018 maximum with $U\sim 6.2$\,mag was the highest one after 
the major 1977-1984 outburst ($U\sim$5\,mag). 
The different time of the maximum in $R_{\rm C}$ and $UBV$ 
filters can be caused by the presence of a $\sim95$ day period 
easily recognized in $B, V$ and $R_{\rm C}$ LCs 
\citep[see][]{2010PASP..122...12W}. 
Further, we observed a sharp minimum at $JD\sim2\,458\,307$ 
($U\sim8.8$\,mag) and then another decrease by $\sim1$\,mag 
until our last observation (see Fig.~\ref{ch_close}). 
%
%
%
\subsection{CI~Cyg}
\label{ss:cicyg}

\begin{figure}[p!t]
\centerline{
\includegraphics[angle=-90, width=1.0\textwidth,clip=]{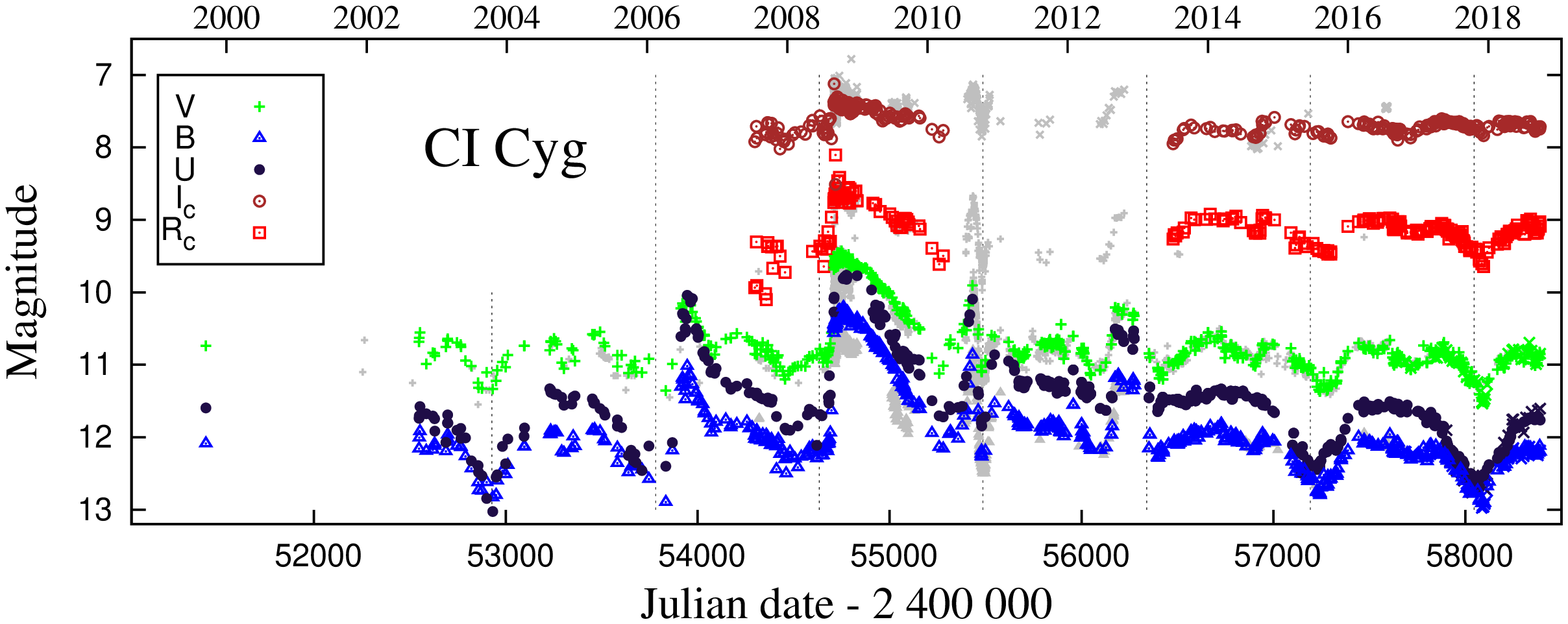}}
\caption{
As in Fig.~\ref{egand}, but for CI~Cyg. The timing of eclipses 
is given by the ephemeris of \cite{2012AN....333..242S}, 
$JD_{\rm Ecl.} = 2\,441\,838.8(\pm1.3) + (852.98\pm0.15)\times E$.
Our observations are displayed with those of 
\cite{2007AN....328..909S,2012AN....333..242S}, 
\cite{2009MNRAS.399.2139S} and from \textsl{AAVSO} (grey). 
}
\label{cicyg}
\end{figure}

\begin{figure}
\centerline{
\includegraphics[angle=-90, width=1.0\textwidth,clip=]{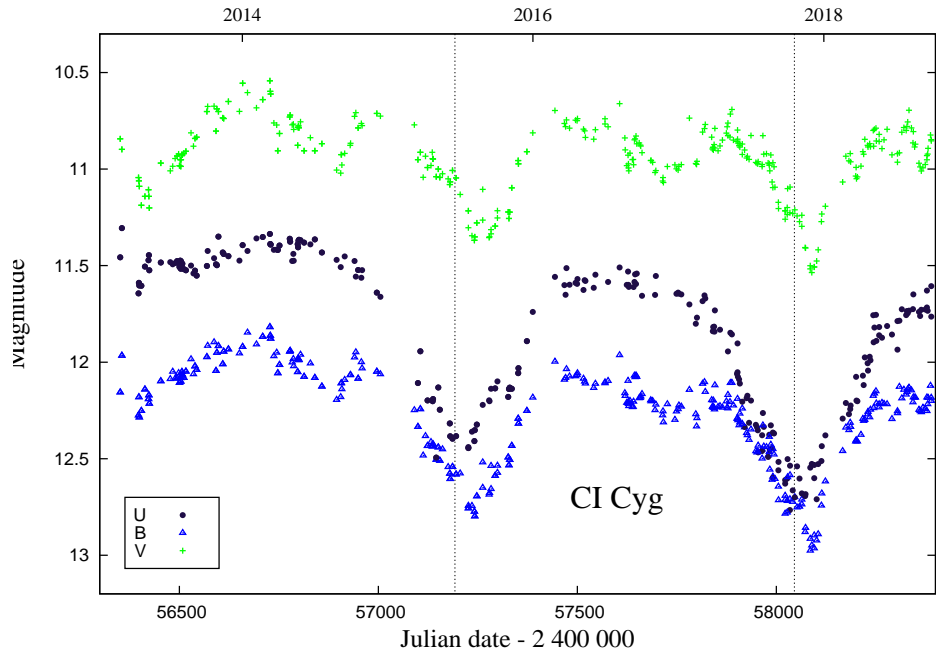}}
\caption{
During the last two orbital cycles, wave-like orbitally-related 
variation developed in the LC of CI~Cyg -- a signature of 
quiescent phases of symbiotic stars (see Sect.~\ref{intr}). 
Secondary minima around $\varphi\sim0.6$ are indicated in $B$ 
and $V$. In the $V$ LC, a variation with a period of $\sim73$ 
days can be found. Vertical lines as in Fig.~\ref{cicyg}. 
}
\label{cicyg_ecl}
\end{figure}

Since September 2008, when CI~Cyg underwent its last major outburst, 
the return to the quiescent phase was interrupted only by two eruptions 
of lower magnitude (Fig.~\ref{cicyg}). The first one, disrupted by 
the eclipse, was described by \cite{2012AN....333..242S}. 
The second one with two maxima around $U = 10.5$\,mag was observed 
during August-September, 2012. 
Since then, CI~Cyg entered a quiescent phase showing the typical 
brightness variation strongly modulated by the orbital motion in 
all filters. A secondary minimum around the $\varphi = 0.6$ was 
recognizable in the $BVR_{\rm C}I_{\rm C}$ LCs 
(see Figs.~\ref{cicyg} and \ref{cicyg_ecl}). 
%
%

According to the basic composition of symbiotic binaries 
(see Sect.~\ref{intr}), the amplitude of such variability is 
wavelength dependent with 
$\Delta U>\Delta B>\Delta V>\Delta R_{\rm C}>\Delta I_{\rm C}$. 
Fig.~\ref{cicyg_ecl} shows that times of the recent wave-like 
minima are shifted from those predicted by the ephemeris for 
eclipses. In addition, the shift seems to be a function of 
the colour. 
For the 2015 minimum, we determined the light minima by 
a simple parabolic fit to Min(B)$\sim $JD~2\,457\,237.2 
and Min(V)$\sim $JD~2\,457\,253.8. For the 2017 minimum, we 
found its timings at Min(B)$\sim$ JD~2\,458\,074.4 and 
Min(V)$\sim $JD~2\,458\,088.4. During the last orbital 
cycle, the LC in all filters is modulated by a light variation 
with the period of $\sim 73$ days, which is most pronounced 
in the $V$ filter (Fig.~\ref{cicyg_ecl}). 

Since 2011 until our last observations, the overall brightness 
of CI~Cyg has been gradually decreasing with a rate of about 
0.17\,mag per year. 
\begin{figure}[p!t]
\centerline{
\includegraphics[angle=-90, width=1.0\textwidth,clip=]{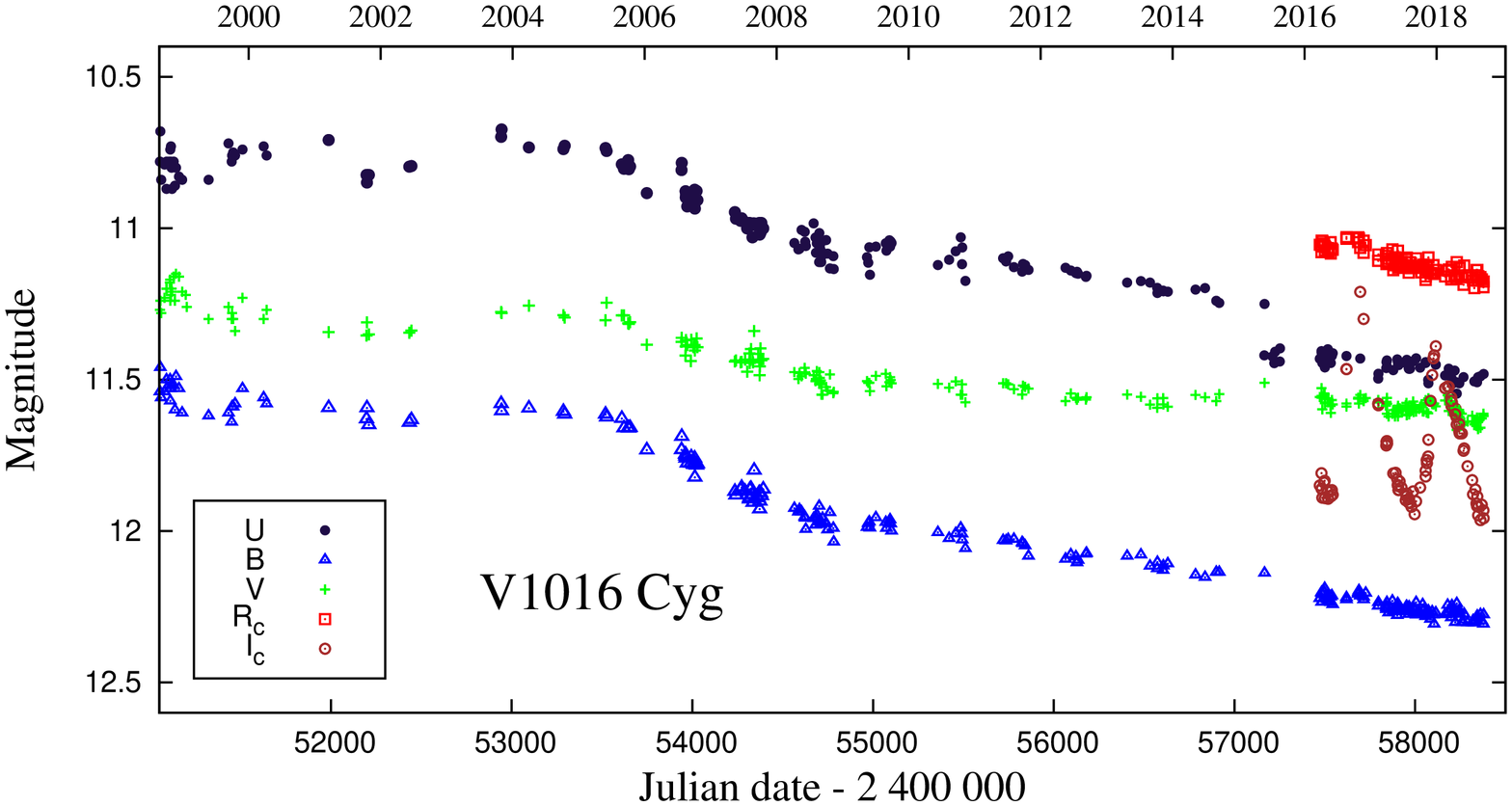}}
\caption{
As in Fig.~\ref{egand}, but for V1016~Cyg. Here, our new 
observations are displayed with those published by 
\cite{2000CoSka..30...99P} and 
\cite{2008AstL...34..474A,2016BaltA..25...35A,2016yCat..90410665A}. 
}
\label{v1016}
\end{figure}
\subsection{V1016~Cyg}
\label{ss:v1016}
\begin{figure}
\centerline{
\includegraphics[angle=-90, width=1.0\textwidth,clip=]{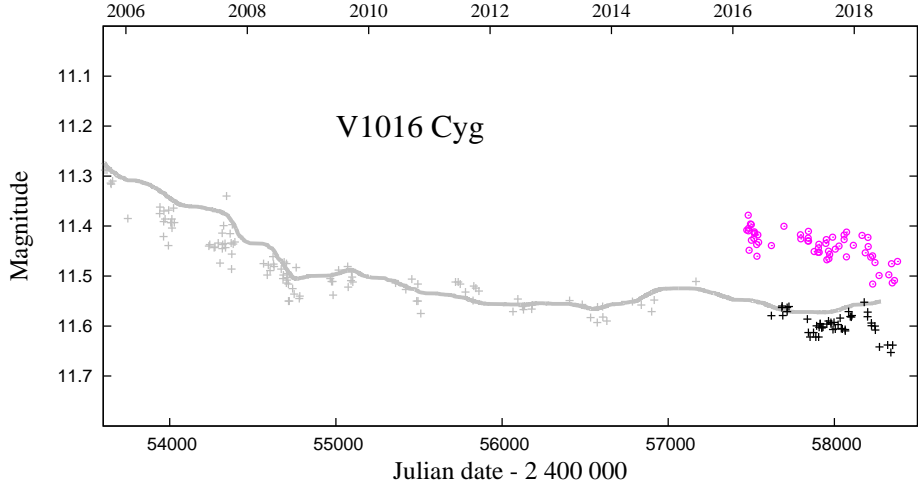}}
\caption{
The $V$ LC of V1016~Cyg using different instruments. Gray crosses 
are data from 
\cite{2008AstL...34..474A,
2015AstL...41..613A,
2016BaltA..25...35A,
2016yCat..90410665A}, 
black crosses are from the private station (M.V.) and G1 
pavilion, while the magenta circles are data obtained in 
the G2 pavilion (see Sect.~\ref{s:obs}). The grey line 
represents smoothed LC of visual estimates from the 
\textsl{AAVSO} database shifted by +0.33\,mag. 
}
\label{v1016_shift}
\end{figure}
\begin{figure}
\centerline{
\includegraphics[angle=-90, width=1.0\textwidth,clip=]
                {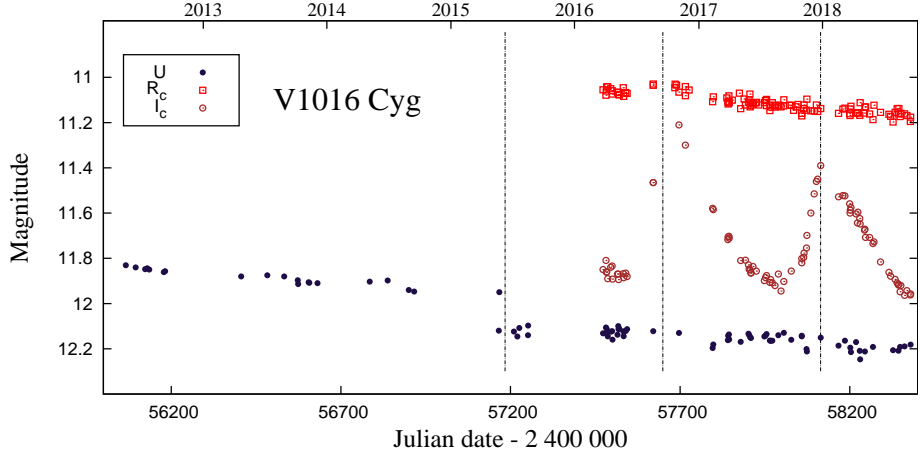}}
\caption{A sudden decrease in the $U$ brightness of V1016~Cyg 
(around 2015.4) could be connected with pulsations of the Mira 
variable. The vertical lines correspond to times of the maximum 
light in the $I_{\rm C}$ filter for the period of 465 days. 
First six points after the decrease are adapted from Fig.~1 
of \cite{2016BaltA..25...35A}. Their uncertainties are 
$<$0.01\,mag and $\pm 10$ days in the scale of magnitudes 
and JD, respectively. 
}
\label{v1016s}
\end{figure}
\begin{figure}
\centerline{
\includegraphics[angle=-90, width=1.0\textwidth,clip=]{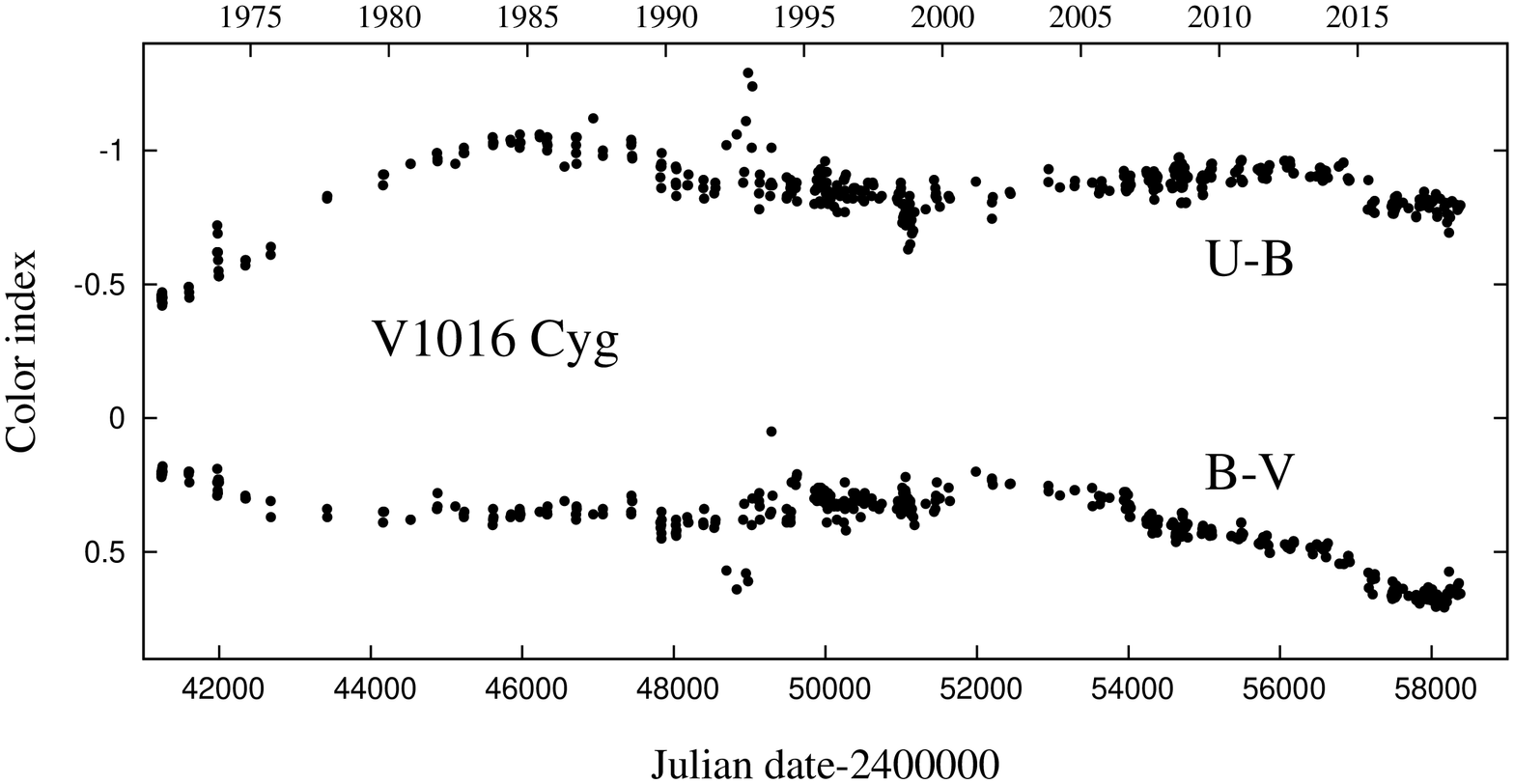}}
\caption{
Evolution of colour indices of V1016~Cyg. 
Data sources as in Fig.~\ref{v1016}. 
}
\label{v1016_CI}
\end{figure}

V1016~Cyg underwent a nova-like outburst in 1964 
\citep[][]{1965IAUC.1916....0M} 
when increased its brightness by $\sim$6\,mag and reached the 
maximum in 1971. Since then the brightness of the system slowly 
decreases most of the time. 
V1016~Cyg is a D-type symbiotic star, with the oxygen-rich cool giant 
of the Mira-type variable 
\citep[][]{1974ApJ...188...95H,1992MNRAS.258...95S}. 
\cite{2015AstL...41..613A} refined the pulsation period of 
the giant to $P = 465\pm5$ days with the minimum at 
$JD_{\rm Min} = (2\,444\,003\pm465)\times E$. 

Here, we present $UBVR_{\rm C}I_{\rm C}$ photometric observations 
of V1016~Cyg obtained after March 2016 (Fig.~\ref{v1016}). 
Pulsations of the giant are clearly apparent in the $I_{\rm C}$ 
filter. The last maximum was observed at 
$JD_{\rm Max}\sim$2\,458\,114 (Fig.~\ref{v1016s}). 
We noticed certain shifts between the data observed in the same 
passband, but using a different telescope--camera--filter 
composition (Fig.~\ref{v1016_shift}). The shifts are as high as 
$\sim0.15$\,mag in $B$ and $V$ filter. This is probably caused 
by the effect usually observed in the LCs of novae during the 
`nebular phase'. The observed flux within a filter is dominated 
by nebular lines rather than by the underlying continuum. 
When using the broadband photometric filters, some prominent 
emission lines can be located close to the edge of the filter 
sensitivity. Critical emission lines for this case are H$\beta$, 
O[III]~4959, 5007\,\AA, H$\alpha$ and Pa$\alpha$, which significantly 
contribute to the fluxes within $B$, $V$, $R_{\rm C}$ and $I_{\rm C}$
filters. Hence, a small difference in the response curve of the 
filter used can result in some shifts in the observed magnitude 
\citep[e.g.][]{2015NewA...40...28M,2015AstL...41..613A}. 
%
Figure~\ref{v1016_shift} demonstrates the effect of strong 
nebular lines located near to the short-wavelength edge of 
the $V$ filter for V1016~Cyg. 
%

In May 2015, \cite{2016BaltA..25...35A,2016yCat..90410665A} 
observed a sudden decrease in the $U$ star's brightness by 
about 0.17\,mag. Our new observations confirm this brightness 
drop (see Fig.~\ref{v1016}). At the same time 
$(U-B)$ and $(B-V)$ colours became redder (Fig.~\ref{v1016_CI}). 
This reddening could be partly caused by the variable circumstellar 
extinction due to the dust produced by the Mira variable and/or 
by the change of the relative intensity of O[III] 4959 and 5007\,\AA\ 
emission lines, because of their different contribution to 
the $B$ and $V$ magnitude \citep[see][]{2015AstL...41..613A}. 
It is of interest to note that the decline in the $U$ and $B$ 
brightness during 2016 occurred around the time of the maximum 
brightness of the Mira variable for the 465 days pulsation period 
(at $JD_{\rm Max}\sim$2\,457\,184, see Fig.~\ref{v1016s}). 
%
%
\subsection{V1329~Cyg}
\label{ss:v1329}
V1329~Cyg was originally an inactive star of about 15\,mag 
displaying $\sim$2\,mag deep eclipses 
\citep[][]{1974AJ.....79...47S,1997A&A...324..606S}. 
In 1964, V1329~Cyg underwent a nova-like eruption when 
brightened by $\Delta m_{\rm pg}\sim$3\,mag, peaked around 
1968 at $B\sim 12.0$\,mag and developed a wave-like 
orbitally-related variation in its LC 
\citep[][]{1969IBVS..384....1K,
1973IBVS..762....1A,1979BAICz..30..308G}. 
This type of photometric variability -- characteristic 
for symbiotic binaries during quiescent phases -- dominates 
the LC of V1329~Cyg to the present 
\citep[e.g.][ and references therein]{2012AN....333..242S}. 

\begin{figure}[p!t]
\centering
\begin{center}
\resizebox{10cm}{!}{\includegraphics[angle=-90]{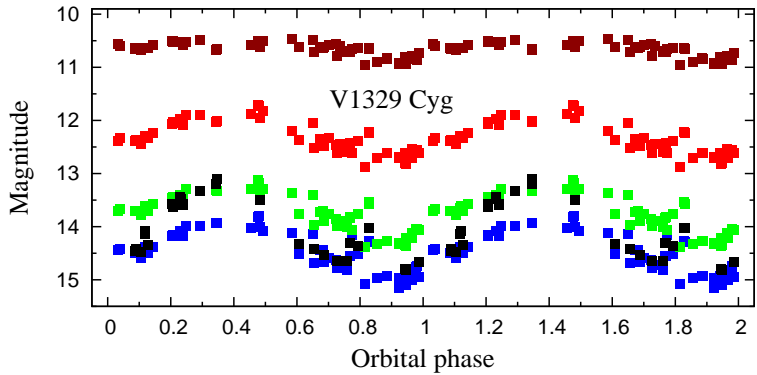}}
\caption{
Phase diagram of our $UBVR_{\rm C}I_{\rm C}$ magnitudes of V1329~Cyg. 
We used the ephemeris for the pre-outburst minima (eclipses) as 
derived by \cite{1997A&A...324..606S}, 
$JD_{\rm Min} = 2\,427\,687(\pm 20) + (958.0\pm 1.8)\times E$. 
Denotation of points as in Fig.~\ref{egand}. 
}
\label{fig:v1329}
\end{center}
\end{figure}

Our new observations of V1329~Cyg are shown in 
Fig.~\ref{fig:v1329}, where they are folded as a function 
of the orbital phase using the ephemeris of the pre-outburst 
eclipses derived by \cite{1997A&A...324..606S}. 
The data cover the period from November 19, 2009, to 
December 25, 2017, i.e. $\sim$3.1 orbits. Our measurements 
confirmed the presence of the orbitally-related light variability 
with a decreasing amplitude to longer wavelengths: 
$\Delta U\sim 1.6$, $\Delta B\sim 1.1$, $\Delta V\sim 0.9$, 
$\Delta R_{\rm C}\sim 0.7$ and $\Delta I_{\rm C}\sim 0.4$\,mag. 
Such the dependence is the result of the superposition of 
the variable contribution from the nebula and a constant 
light from the red 
giant, whose contribution increases to longer wavelengths. 
Figure~\ref{fig:v1329} also shows that the light minimum 
precedes the inferior conjunction of the giant by 
$\sim 0.1\times P_{\rm orb}$. 
This effect is caused by a prolate structure of the symbiotic 
nebula, whose optically thick part is asymmetrically located 
with respect to the binary axis during quiescent phases of 
symbiotic stars \citep[see][ in detail]{1998A&A...338..599S}. 
%
%

\subsection{AG~Dra}
\label{ss:agdra}

AG~Draconis is a well studied bright symbiotic star. 
It is a yellow symbiotic binary comprising a K2\,{\small III} 
giant \citep[][]{1999A&AS..137..473M} and a WD accreting from 
the giant's wind on a 549-day orbit \citep[][]{2000AJ....120.3255F}. 
There are no signs of eclipses neither in the optical nor far-UV. 
Based on spectropolarimetric observations, 
\cite{1997A&A...321..791S} derived the orbital inclination 
$i = 60{\degr}\pm 8{\degr}.2$. 
The system undergoes occasional eruptions, lasting about 
1-2 orbital cycles. The star’s brightness abruptly increases 
($\Delta U\sim$2, $\Delta V\sim$1\,mag) showing multiple 
maxima separated approximately by 400 days, while during 
quiescent phase the LC shows typical wave-like orbitally-related 
modulation 
\citep[e.g.][ and references therein]{1979IBVS.1611....1M,
1998A&A...338..599S,2014MNRAS.443.1103H,2016MNRAS.456.2558L} 
The historical LC was analyzed by \cite{2012MNRAS.422.2648F}. 
They confirmed the $\sim550$ days orbital period and 
$\sim355$-day pulsation period of the K-type giant, originally 
proposed by \cite{1999A&A...348..533G}. 
Moreover, they reported also $\sim373$ days cycle between 
the outbursts and suggested $\sim1160$-day rotation period 
of the giant. 
\begin{figure}[p!t]
\centerline{
\includegraphics[angle=-90, width=1.0\textwidth,clip=]{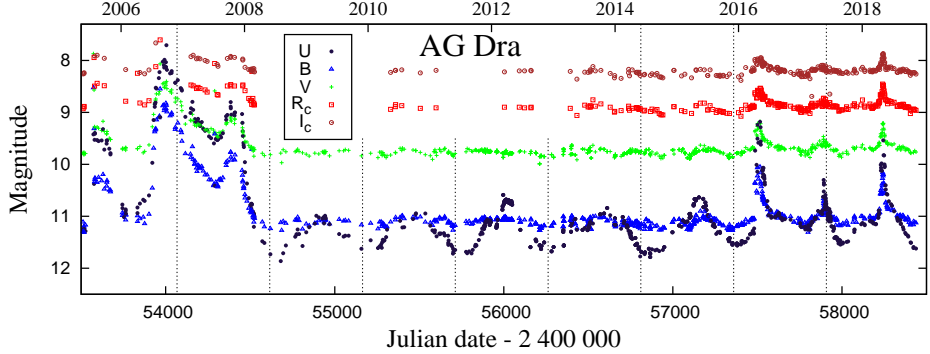}}
\caption{
As in Fig.~\ref{egand}, but for AG~Dra. Vertical lines represent 
inferior conjunctions of the giant according to the ephemeris of 
\cite{2000AJ....120.3255F}, 
$JD_{\rm sp. conj.} = 2\,450\,775.3(\pm4.1) + (548.65\pm0.97)\times E$. 
Our new observations are displayed with the data published by 
\cite{2012AN....333..242S} and \cite{2009PASP..121.1070M}. 
}
\label{agdra}
\end{figure}
\begin{figure}[p!t]
\centerline{
\includegraphics[angle=-90, width=0.9\textwidth,clip=]{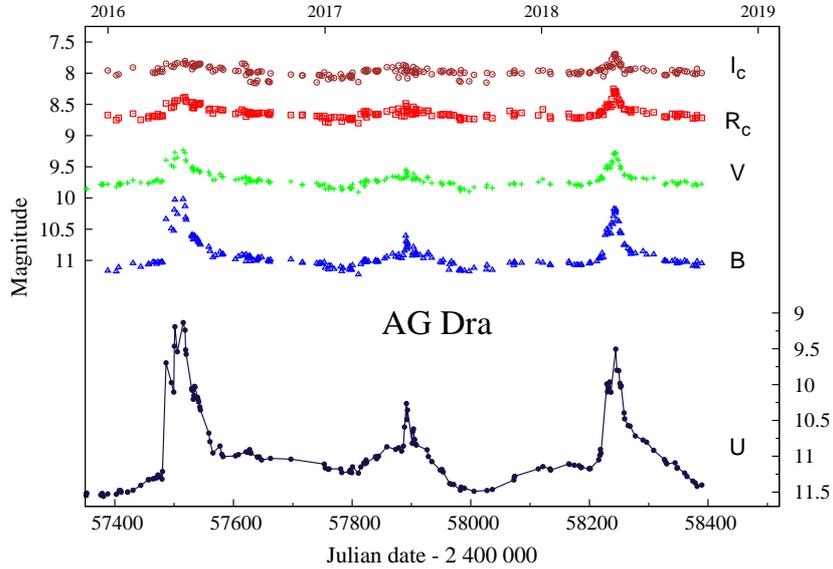}}
\caption{
Recent three outbursts of AG~Dra. Note the multi-peaked structure 
for the 2016 one. 
}
\label{agdra_hot}
\end{figure}
\begin{figure}
\centerline{
\includegraphics[angle=-90, width=0.9\textwidth,clip=]{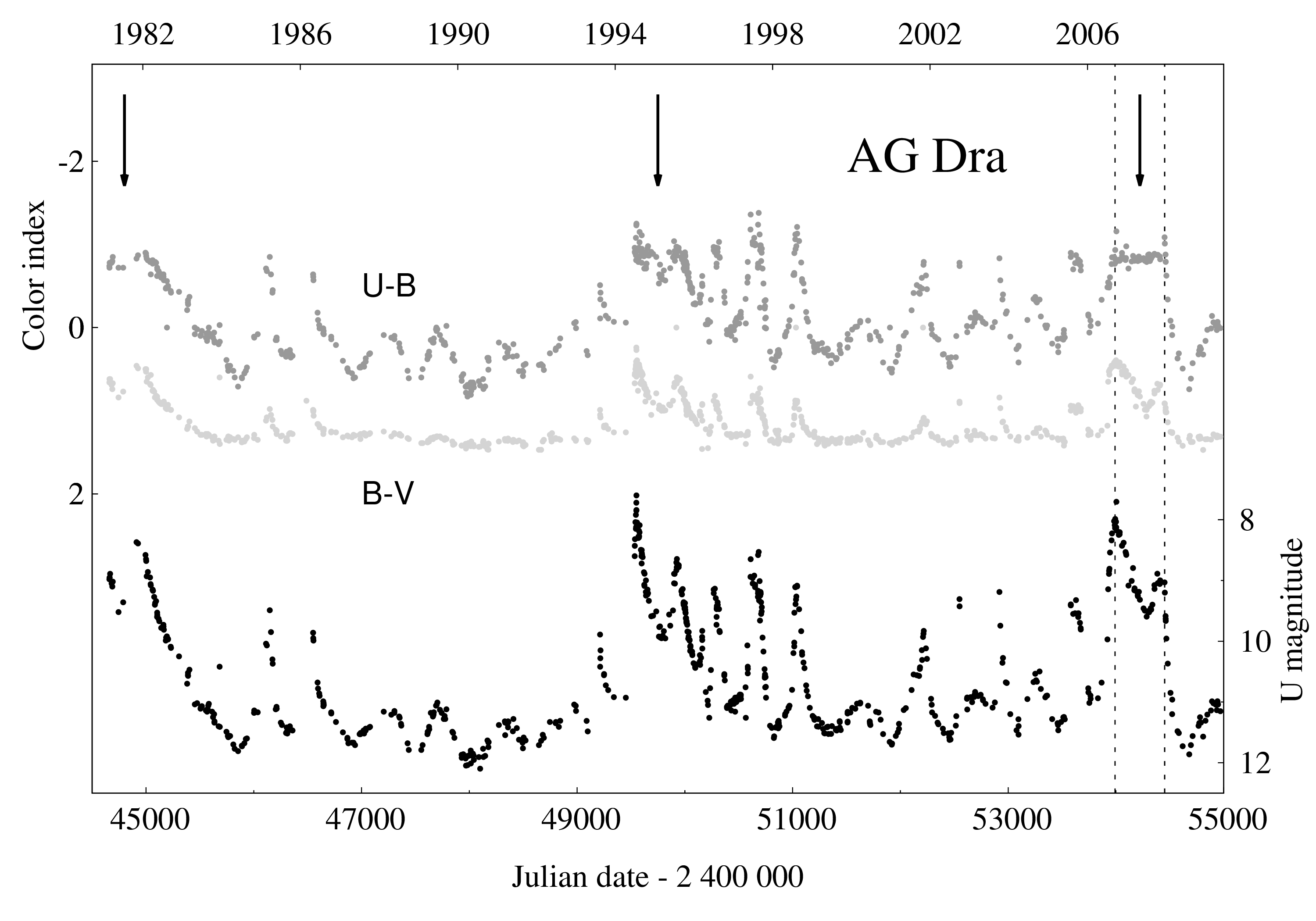}}
\caption{
The $(U-B)$ colour shows a plateau during the cool 2006-07 
outburst of AG~Dra. Vertical lines bounds its duration. Also 
during previous major outbursts (1981-82, 1994-95) the index 
was relatively stable (arrows). 
}
\label{agdra_plat}
\end{figure}

\cite{1999A&A...347..478G} identified cool and hot outbursts 
differing in their Zanstra temperatures and the LC profile. 
The former are more pronounced ($\Delta U\sim3$\,mag) lasting 
for 1--2 years, while the latter are weaker ($\Delta U\sim 1-2$\,mag), 
single brightenings lasting for weeks to months. 
The last cool outburst was observed during 2006--2007. 
The following quiescent phase was interrupted with two small 
brightenings in March 2012 and May 2015 ($U\sim10.6$\,mag; 
see Fig.~\ref{agdra}). Further, three bright bursts were observed 
in April 2016 ($U\sim9.1$\,mag), in May 2017 ($U\sim10.3$\,mag) 
and in May 2018 ($U\sim9.5$\,mag). A multi-peaked profile is 
apparent especially during the 2016 outburst, where it can be 
recognized in $UBVR_{\rm C}$ filters (Fig.~\ref{agdra_hot}).

It is of interest to note that the $(U-B)$ colour practically 
did not change during the last 2007 major outburst 
(see Fig.~\ref{agdra_plat}). The plateau part of the LC occurred 
just between its two maxima. However, the LC profile of the 
$(B-V)$ index resembles that of individual LCs. This plateau 
was not so evident during the 1994 major outburst, while that 
during the 1981-82 outburst was better comparable with 
the recent one. 
The plateau phases during major outbursts of AG~Dra are probably 
caused by a strong dominance of the nebular radiation within 
the $U$ and $B$ passbands \citep[e.g.][]{2005A&A...440..995S}. 
Then the star's brightness decline here is given by that in 
the nebular emission, which results in more or less constant 
value of the $(U-B)$ index. However, the brightness decrease 
in, e.g., $V$ filter is slower than in a bluer one, because of 
a significant contribution from the giant in $V$, and thus the 
$B-V$ index is more similar to individual LC profile 
(see Fig.~\ref{agdra_plat}). 
%

\subsection{Draco~C1}
\label{ss:dracoc1}
The symbiotic star Draco C1 is located in the Draco dwarf 
spheroidal galaxy (DdSg). Its symbiotic nature was discovered by 
\cite{1982ApJ...254..507A} within a grating prism survey of 
the DdSg. The spectrum was characterized by strong emission 
lines superposed on a carbon star continuum. 
Draco~C1 has been studied photometrically for a long time 
\citep[][]{1991IBVS.3605....1M,
           2007AN....328..909S,
           2008AJ....136.1921K,
           2008BaltA..17..293H,
           2008IBVS.5855....1M,
           2012AN....333..242S}. 
Long-term photometry showed a variable LC with 
$\Delta V\sim$0.1\,mag and periodicity around one year, 
which could be attributed to the orbitally-related variation 
of quiescent symbiotic stars. 

Our new $BVR_{\rm C}I_{\rm C}$ measurements of Draco~C1 
cover the period between November 6, 2010, and August 6, 2016. 
%

\subsection{RS~Oph}
\label{ss:rsoph}

\begin{figure}[p!t]
\centerline{
\includegraphics[angle=-90, width=1.0\textwidth,clip=]{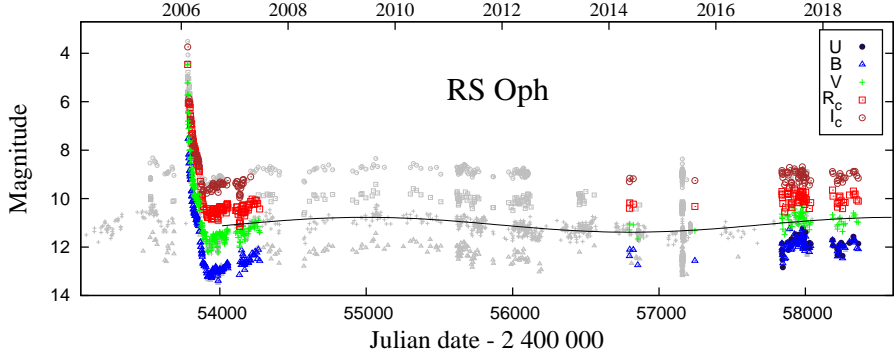}}
\caption{
As in Fig.~\ref{egand}, but for RS~Oph. Our new observations 
are displayed with the data of \cite{2008ASPC..401..206H} 
(2006 outburst) and from the \textsl{AAVSO} database (grey). 
The solid line represents a sine function with amplitude 
$\Delta V\sim0.3$\,mag and a period of $\sim$9.4 years. 
}
\label{rsoph}
\end{figure}

\begin{figure}[p!t]
\centering
\resizebox{10cm}{!}{\includegraphics[angle=-90]{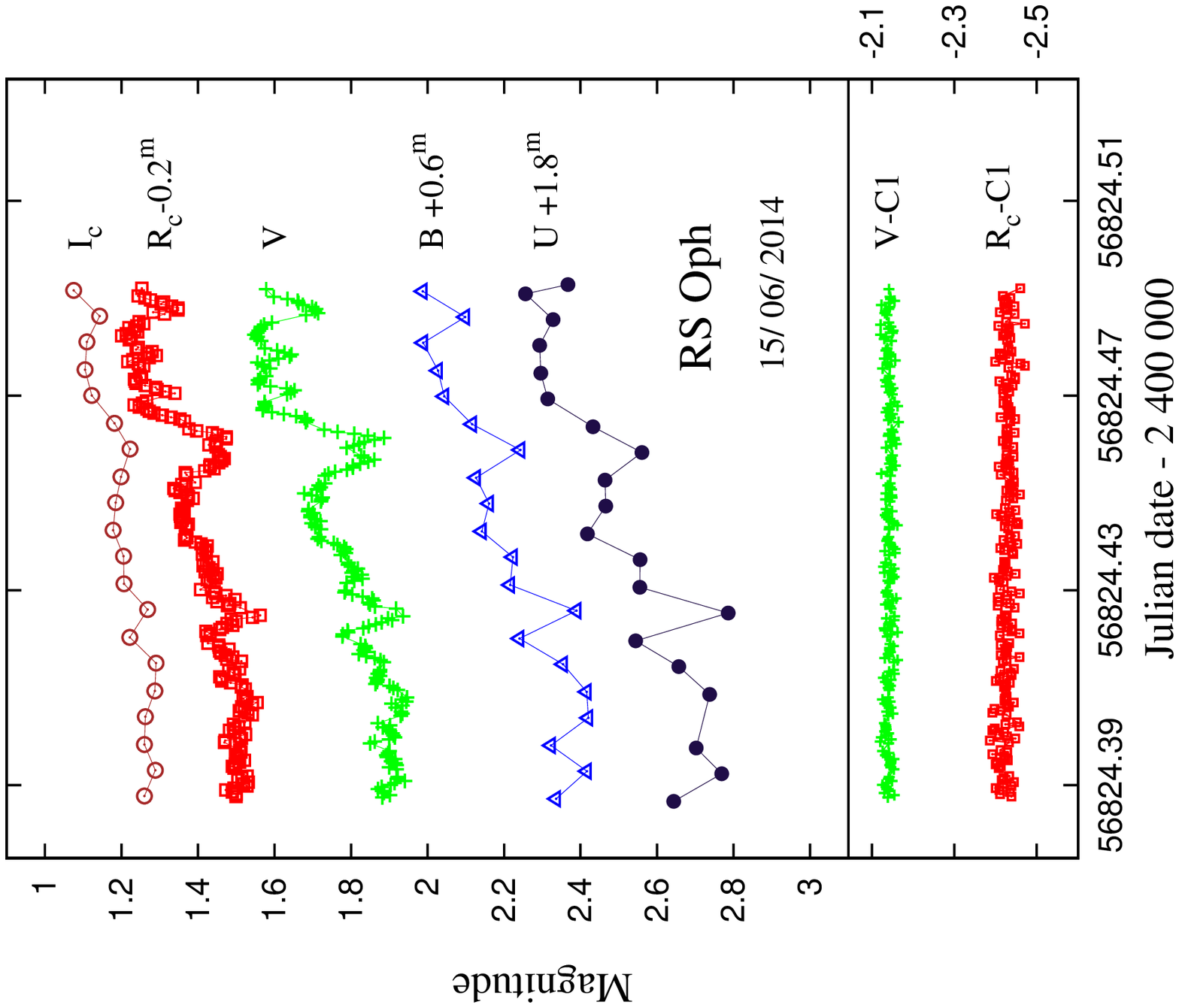}}\vspace*{5mm}
\resizebox{\hsize}{!}{\includegraphics[angle=-90]{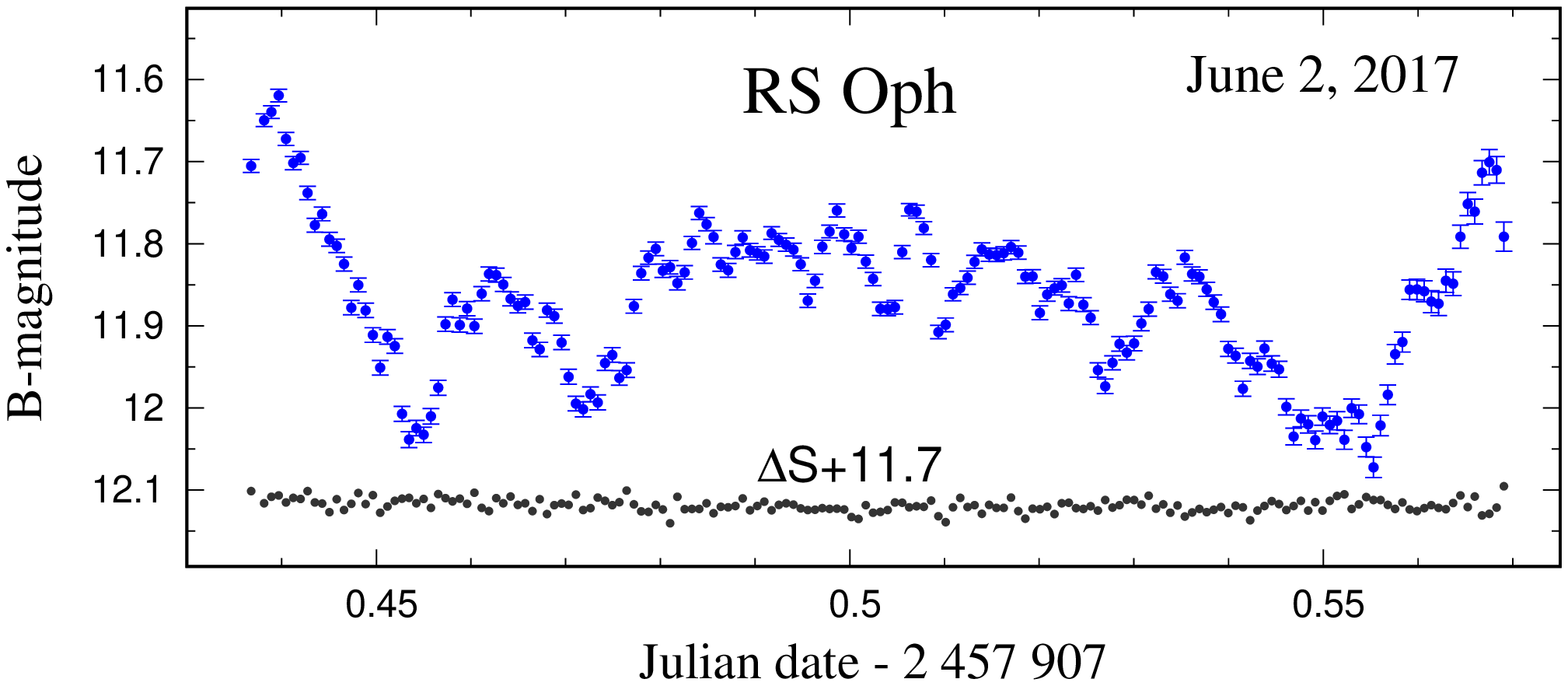}}
\caption{
Examples of rapid photometric variability of RS~Oph. Top panel 
shows variations in $UBVR_{\rm C}I_{\rm C}$ filters obtained in 
G2 pavilion (point (ii) of Sect.~\ref{s:obs}), while the bottom 
panel shows high-cadence observations in $B$ acquired at the 
private station (point (iii) of Sect.~\ref{s:obs}). 
Here, we used the star UCAC4-417-071413, ($V$=9.307\,mag, $B-V$=1.236) 
as the comparison and UCAC4-416-072918 as the check star. 
$\Delta$S denotes their difference in $B$. 
%
}
\label{rsoph_flick}
\end{figure}

RS~Oph is a symbiotic recurrent nova with the orbital period 
$P_{\rm orb}=453.6 \pm 0.4$ days \citep[][]{2009A&A...497..815B}. 
To the present, six eruptions were recorded in 1898, 1933, 1958, 
1967, 1985 and the last one in 2006. 
Additional two outbursts probably happened in 1907 and 1954. 
For the former, only a post-eruption dip was observed on 
archival photographic plates, because of the season gap in 
observations \citep[][]{2004IAUC.8396....2S}. Similarly, the 
latter was suggested by \cite{1993AAS...183.5503O} according 
to the presence of such the dip and the fading tail from the 
eruption, later confirmed by \cite{2011MNRAS.414.2195A}. The 
mean time between the eruptions is $\sim$15 years. After the 
2006 outburst, RS~Oph remains in a quiescent phase, during 
which the LC shows a sine-like variation with the amplitude 
$\Delta V\sim0.3$\,mag and a period of $\sim$9.4 years 
(Fig.~\ref{rsoph}). 
A large scatter of the data is caused by 
the presence of flickering with the amplitude of several tenths 
of a magnitude \citep[e.g.][]{1986A&A...167...91B,
2007MNRAS.379.1557W,2015MNRAS.450.3958Z}. 
Our measurements of the rapid light variability of RS~Oph 
are showed in Fig.~\ref{rsoph_flick}. It is observed in all 
$UBVR_{\rm C}I_{\rm C}$ filters having the amplitude decreasing 
with increasing wavelength, because of increasing constant 
contribution from the giant in longer wavelengths. 
%
%
\subsection{AR~Pav}
\label{ss:arpav}
AR~Pav is eclipsing symbiotic binary with the orbital 
period of 605 days \citep[][]{1937AnHar.105..491M}. 
It consists of an M5\,III giant with a mass of $\sim$2\,\mo\ 
\citep[][]{1999A&AS..137..473M}. According to the observed 
variations in the UV/optical continuum 
\citep[e.g.][]{2001A&A...366..972S,2000MNRAS.311..225S}, 
the hot component is highly variable in brightness, size and 
the geometry as suggested by very different profiles of 
minima -- eclipses \citep[][]{1994A&A...287..829B,
2000MNRAS.311..225S,2001IBVS.5195....1S}. 

\begin{figure}[p!t]
\centerline{
\includegraphics[angle=-90, width=0.9\textwidth,clip=]{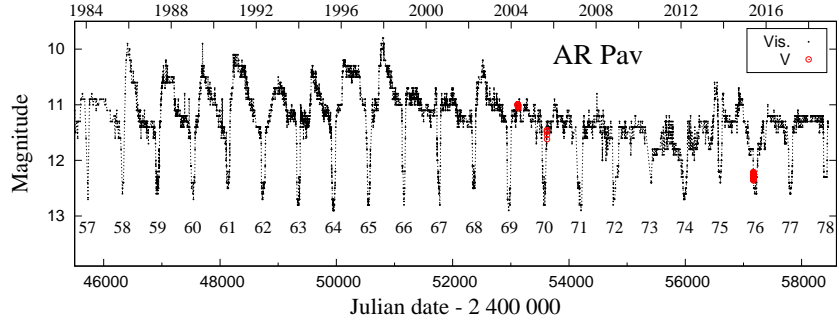}}
\caption{The visual LC of AR~Pav made by two observers (AJ and RS). 
Our new observations (see Sect.~\ref{s:obs}) are displayed with 
the $V$ data from the \textsl{AAVSO} database (Rolf Carstens) and 
those published by \cite{2004CoSka..34...45S,2007AN....328..909S}. 
}
\label{arpav}
\end{figure}

\begin{figure}
\centerline{
\includegraphics[angle=-90, width=0.9\textwidth,clip=]{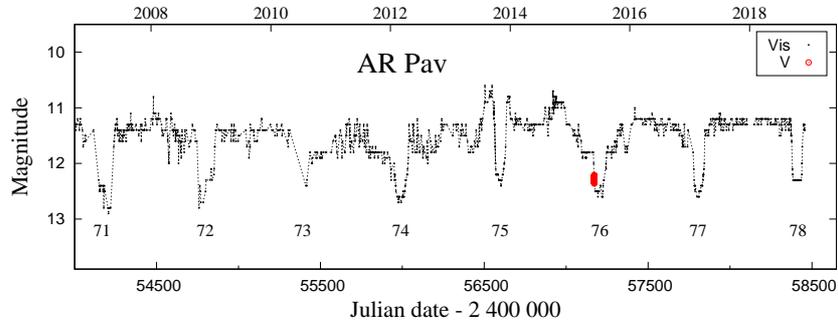}}
\caption{
The recent low stage of AR~Pav between the epoch 71 and 78. 
}
\label{ar_close}
\end{figure}

Photographic magnitudes of AR~Pav were recorded on Harvard 
chart plates since 1889 \citep[][]{1937AnHar.105..491M}. 
Monitoring this star during 1889-1937, the authors classified 
it as an eclipsing binary of the P~Cyg type with an orbital 
period of 605 days. Comparing consecutive refinements of 
the orbital period 
\citep[see][]{1974MNRAS.167..635A,1994A&A...287..829B}, 
complemented with new, 1985-1999, observations, 
\cite{2000MNRAS.311..225S} found a decreasing trend of 
the orbital period. The following observations 
\citep[see][]{2001IBVS.5195....1S,2004CoSka..34...45S} 
supported this result until the period felt to the value 
of $602.8\pm 0.3$ days as determined from the minima (66), 
(67) and (68) (Figs.~\ref{arpav} and \ref{ar_close}). 
The period between Min(69) and Min(70), however, raised up 
to $607.3\pm2.1$ days \citep[][]{2007AN....328..909S}. 
Authors ascribed this difference to a strongly variable size, 
geometry and radiation of the eclipsing object, which affects 
the variation in the minima depth and the overall profile of 
the eclipse. 
\begin{table}
\begin{center}
\caption{The timing of eclipses in the historical LC of AR~Pav}
\label{tpav}
\begin{tabular}{rrcrrc}
\hline
\hline
~~~Epoch&$JD_{\rm Min}$~~~~~&Ref.$^{\star}$&~~~Epoch&$JD_{\rm Min}$~~~~~
                         &Ref.$^{\star}$ \\
\hline
4  & 2413678.4$\pm$1.9 & 1  &    62 & 2448742.6$\pm$0.6& 1  \\
6  &   14881.3$\pm$1.5 & 1  &    63 &   49347.8$\pm$0.6& 1  \\
9  &   16705.3$\pm$2.3 & 1  &    64 &   49951.8$\pm$0.5& 1  \\
12 &   18519.0$\pm$3.8 & 1  &    65 &   50555.5$\pm$0.8& 1  \\
15 &   20333.4$\pm$3.6 & 1  &    66 &   51158.9$\pm$0.7& 2  \\
21 &   23960.6$\pm$2.0 & 1  &    67 &   51762.8$\pm$0.7& 2  \\
24 &   25776.2$\pm$3.0 & 1  &    68 &   52364.5$\pm$0.7& 3  \\
27 &   27589.5$\pm$1.0 & 1  &    69 &   52966.0$\pm$2.0& 4  \\
47 &   39679.1$\pm$1.7 & 1  &    70 &   53573.3$\pm$0.7& 4  \\
50 &   41492.0~~~~~~~  & 1  &    71 &   54188.2$\pm$4.5& tp \\
56 &   45113.7$\pm$2.0 & 1  &    72 &   54793.2$\pm$1.1& tp \\
57 &   45718.7$\pm$2.3 & 1  &    73 &          -       &    \\
58 &   46321.4$\pm$2.2 & 1  &    74 &   55989.9$\pm$0.6& tp \\
59 &   46924.7$\pm$0.4 & 1  &    75 &   56599.2$\pm$0.3& tp \\
60 &   47531.7$\pm$0.9 & 1  &    76 &   57205.3$\pm$0.9& tp \\
61 &   48138.1$\pm$0.5 & 1  &    77 &   57807.7$\pm$0.8& tp \\
   &                   &    &    78 &   58409.5$\pm$1.0& tp \\
\hline
\hline
\end{tabular}
\end{center}
$^{\star}$ Reference: 1 - \cite{2000MNRAS.311..225S}, 
                      2 - \cite{2001IBVS.5195....1S}, 
                      3 - \cite{2004CoSka..34...45S},
                      4 - \cite{2007AN....328..909S},
                      tp - this paper
\end{table}

\begin{figure}
\centerline{
\includegraphics[angle=-90, width=0.8\textwidth,clip=]{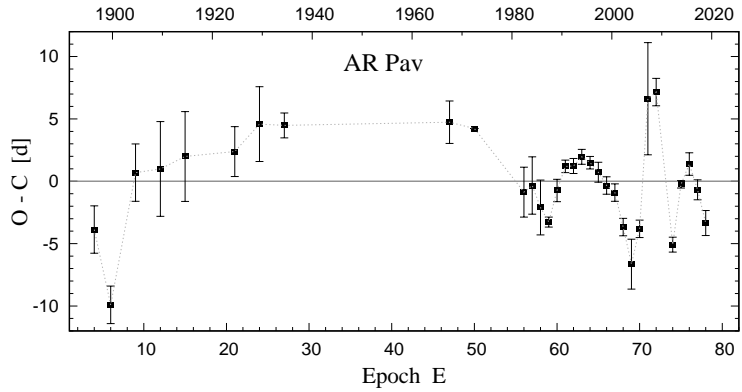}}
\caption{
O--C diagram for the linear ephemeris (\ref{eq:eph}) determined 
by timing of eclipses (Table~\ref{tpav}) in the historical LC 
of AR~Pav. 
}
\label{fig:ar_eph}
\end{figure}

In this paper, we present our (A.J., R.S.) new visual magnitude 
estimates, which are in good agreement with photoelectric 
$V$ magnitudes 
\citep[see][and Fig.~\ref{arpav} here]{2000MNRAS.311..225S,
2001IBVS.5195....1S,2004CoSka..34...45S,2007AN....328..909S}. 
We determined times of new minima between the epoch 71 and 78 
using one or several methods (parabolic fit, tracing paper method, 
Kwee and van Woerden method, sliding integration). Minimum 
at E = 73 was not determined, because of insufficient 
data (Fig.~\ref{ar_close}). Our times of minima (eclipses) 
together with those published in the literature are listed 
in Table~\ref{tpav}. They correspond to linear ephemeris 
\begin{equation}
 JD_{\rm Min} = 2\,411\,264.4(\pm 1.7) + (604.47\pm 0.03)\times E. 
\label{eq:eph}
\end{equation}
Residuals between the observed and predicted times of minima 
(eclipses) are shown in Fig.~\ref{fig:ar_eph}. 
The $O-C$ diagram indicates real systematic variation in the 
minima timing. In general, this can be caused by variable mass 
transfer and mass accretion resulting in variable size and shape 
of the eclipsing object, and thus in a variation of times 
of the light minima. 
%
%
%

During the higher level of the activity, the LC of AR~Pav 
resembles that of AX~Per. However, the maximum light occurs 
usually around the orbital phase $\varphi\sim0.25$ and not during 
the superior conjunction of the giant (Fig.~\ref{outbursts}). 
After 1998, AR~Pav reduced its out-of-eclipse activity to 
only three episodes of the star's brightening; in 2002, 2013 
and 2014 (see Figs.~\ref{arpav} and \ref{ar_close}). 
The flat LC profile between eclipses is observed from 
$\sim$2016.5 to date. 
%
%
\subsection{AG~Peg}
\label{ss:agpeg}

\begin{figure}[p!t]
\centerline{
\includegraphics[angle=-90, width=1.0\textwidth,clip=]{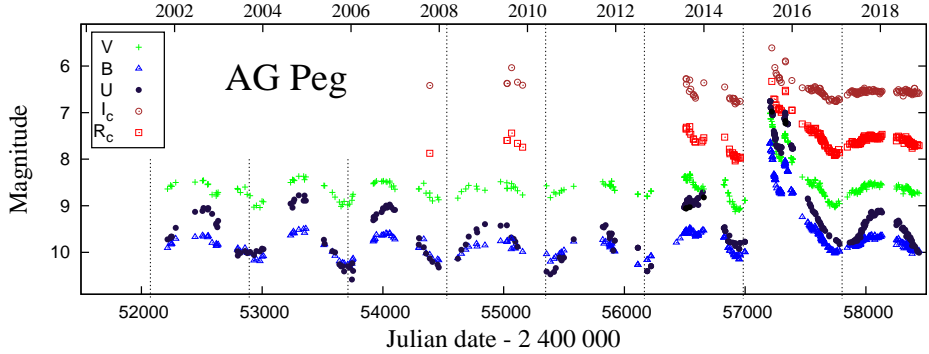}}
\caption{
As in Fig.~\ref{egand}, but for AG~Peg. 
Data to 2011.9 were published by \cite{2012AN....333..242S}. 
Vertical lines represent inferior conjunctions of the giant 
according to ephemeris of \cite{2000AJ....119.1375F}, 
$JD_{\rm sp. conj.} = 2\,447\,165.3(\pm 48) + (818.2\pm 1.6)\times E$.
}
\label{agpeg}
\end{figure}

AG~Peg is a symbiotic binary comprising a M3\,III giant 
\citep[][]{1987AJ.....93..938K} and a WD on an 
818-d orbit \citep[e.g.][]{2000AJ....119.1375F}. 
This system is known as the slowest nova ever recorded 
\citep[][]{1993AJ....106.1573K}. Its nova-like outburst 
began during 1850 \citep[][]{1921AN....213...93L} when 
it rose in brightness from $\sim 9$ to a maximum of 
$\sim 6$\,mag around 1885. Afterwards, AG~Peg followed 
a gradual decline to around 1997 
\citep[][]{1967SvA....11....8B,2001AJ....122..349K}. 
Then AG~Peg entered a quiescent phase with a typical 
orbitally-related wave-like light variations with the 
amplitude $\Delta U\sim1.4$\,mag around the mean of 
9.7\,mag until June 2015, when erupted again showing 
a Z~And-type outburst \citep[Fig.~\ref{agpeg} here, 
Fig.~1 of][and references therein]{2017A&A...604A..48S}. 
%
\begin{figure}
\centerline{
\includegraphics[angle=-90, width=1.0\textwidth,clip=]{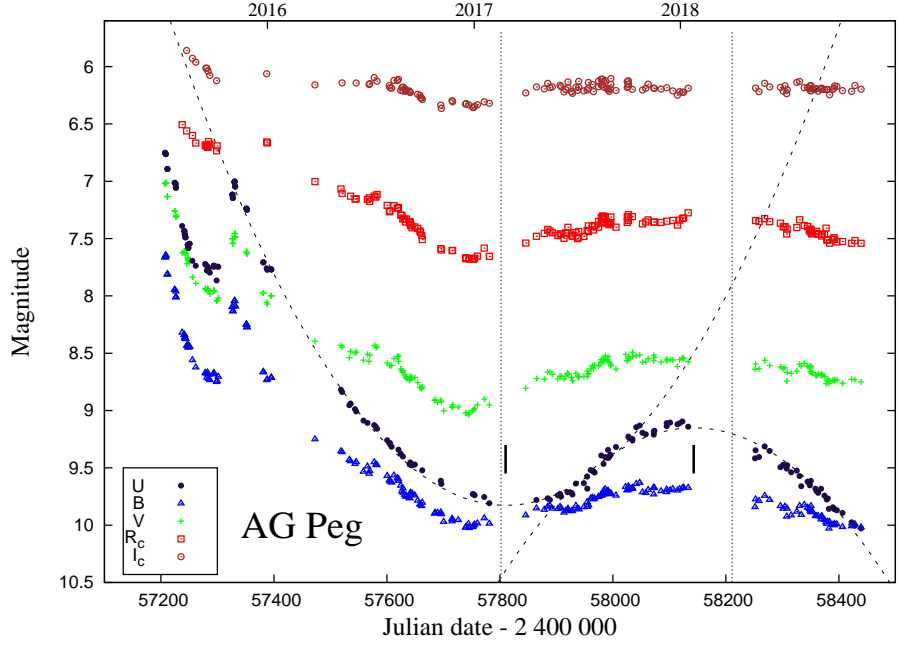}}
\caption{
Multicolour LCs after the 2015 outburst demonstrating transition 
of AG~Peg to quiescent phase. Dashed curves are quadratic 
functions, vertical lines represents orbital phase 0 
(JD~2\,457\,801.9) and 0.5 (JD~2\,458\,211.0) and short 
vertical bars mark corresponding minimum and maximum in $U$.} 
\label{agp_close}
\end{figure}

The 2015 outburst was first reported by \cite{2015AAN...521....1W}. 
According to our multicolour photometry 
\citep[Fig.~\ref{agpeg} here and 
Fig.~2 of][]{2017A&A...604A..48S}, the outburst started 
around June 5, 2015 (JD$\sim2\,457\,177.6\pm3.3$) 
at $B = 9.55\pm 0.07$\,mag and $V = 8.45\pm 0.06$\,mag, reaching 
a maximum around June 30, 2015 (JD$\sim2\,457\,203.5\pm2.0$) at 
$B = 7.68\pm0.05$\,mag and $V = 7.0\pm0.1$\,mag. After 
a gradual decline, a sudden re-brightening occurred on 
October 8, 2015 (JD$\sim2\,457\,304.0\pm0.5$) with a rise 
in $U$, $B$ and $V$ by $\sim$1, $\sim$0.7 and 0.5\,mag, 
respectively. 
In contrast to the first maximum, the secondary one was 
followed by a plateau phase until November 24, 2015, when 
the brightness began to decrease gradually. 

From August 2016 to the end of our observations (December 2018), 
the LCs developed a wave, signalizing that AG~Peg returned to 
a quiescent phase (see Figs.~\ref{agpeg} and \ref{agp_close}). 
The shape of the $U$ LC could be compared to quadratic functions, 
which aid us to determine its extrema. 
The minimum and maximum occurred at $U\sim 9.8$ and $\sim 9.1$
at the orbital phase $\varphi\sim0.01$ and $\sim0.4$, 
respectively. 

The brightness, the short timescale and the multiple peaks of 
the 2015 outburst refer to a different type of outburst 
in comparison with the event in the mid of 19th century. 
It was shown that the recent outburst was of a Z~And-type and 
that AG~Peg obviously entered a new era of its evolution 
\cite[][]{2016MNRAS.462.4435T,2016MNRAS.461.3599R,
          2017A&A...604A..48S}. 
%
%
\subsection{AX~Per}
\label{ss:axper}
\begin{figure}[p!t]
\centerline{
\includegraphics[angle=-90, width=1.0\textwidth,clip=]{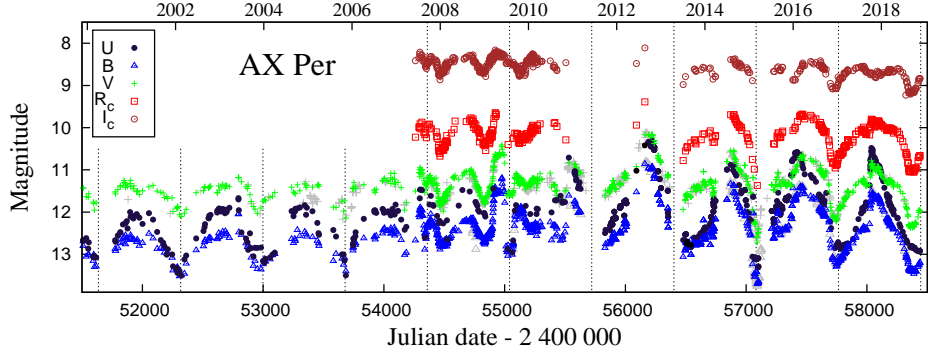}}
\caption{
As in Fig.~\ref{egand}, but for AX~Per. Vertical lines 
represent the timing of eclipses, 
$JD_{\rm Ecl.}=2\,447\,551.26(\pm 0.3)+(680.83\pm 0.11)\times E$
\citep[][]{2011A&A...536A..27S}. New data are complemented 
with those of \cite{2011A&A...536A..27S,2012AN....333..242S} 
and $V$ from \textsl{AAVSO} (grey). 
}
\label{axper}
\end{figure}
\begin{figure}[p!t]
\centerline{
\includegraphics[angle=-90, width=1.0\textwidth,clip=]{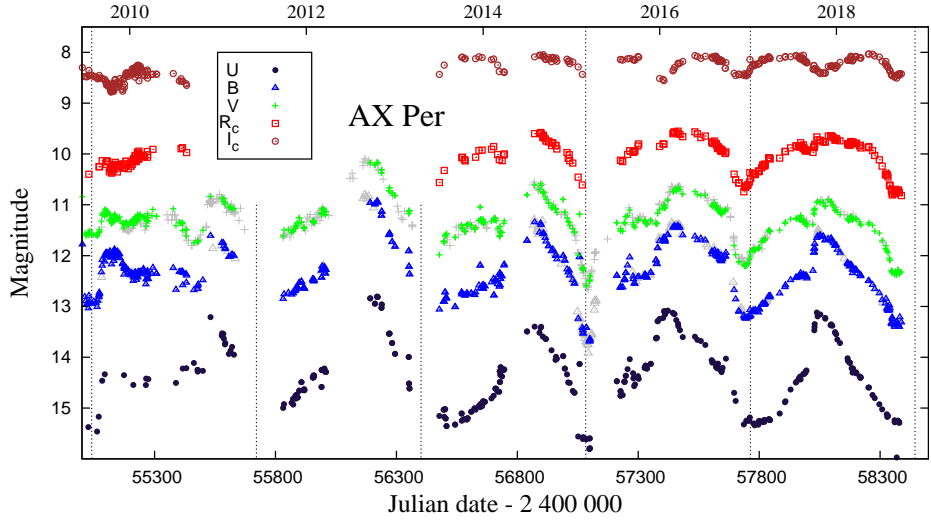}}
\caption{The recent evolution of the AX~Per LCs covering its 
current active stage. The $U$ LC is shifted by +2.5\,mag 
for better visualization. 
}
\label{axper_sep}
\end{figure}
%
\begin{figure}[p!t]
\centerline{
\includegraphics[angle=-90, width=1.0\textwidth,clip=]{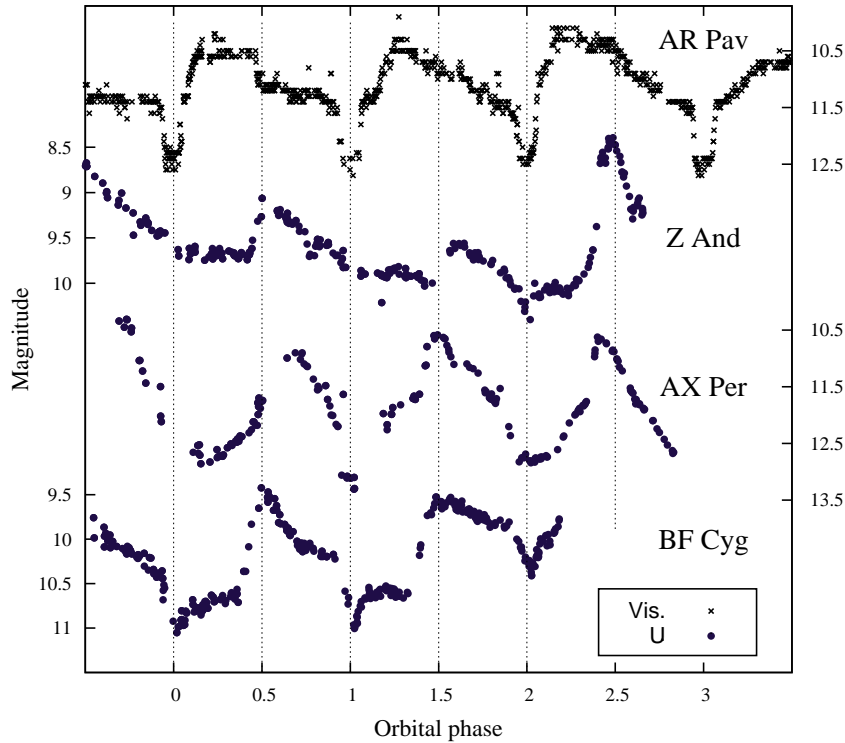}}
\caption{
Comparison of the recent AX~Per LC with those of other active 
symbiotic stars. The light maximum and minimum usually appear 
around $\varphi\sim$0.5 and $\sim$0, respectively, while the 
LC of AR~Pav shows maxima around $\varphi\sim$0.25. 
}
\label{outbursts}
\end{figure}
%
\begin{figure}[p!t]
\centerline{
\includegraphics[angle=-90, width=1.0\textwidth,clip=]{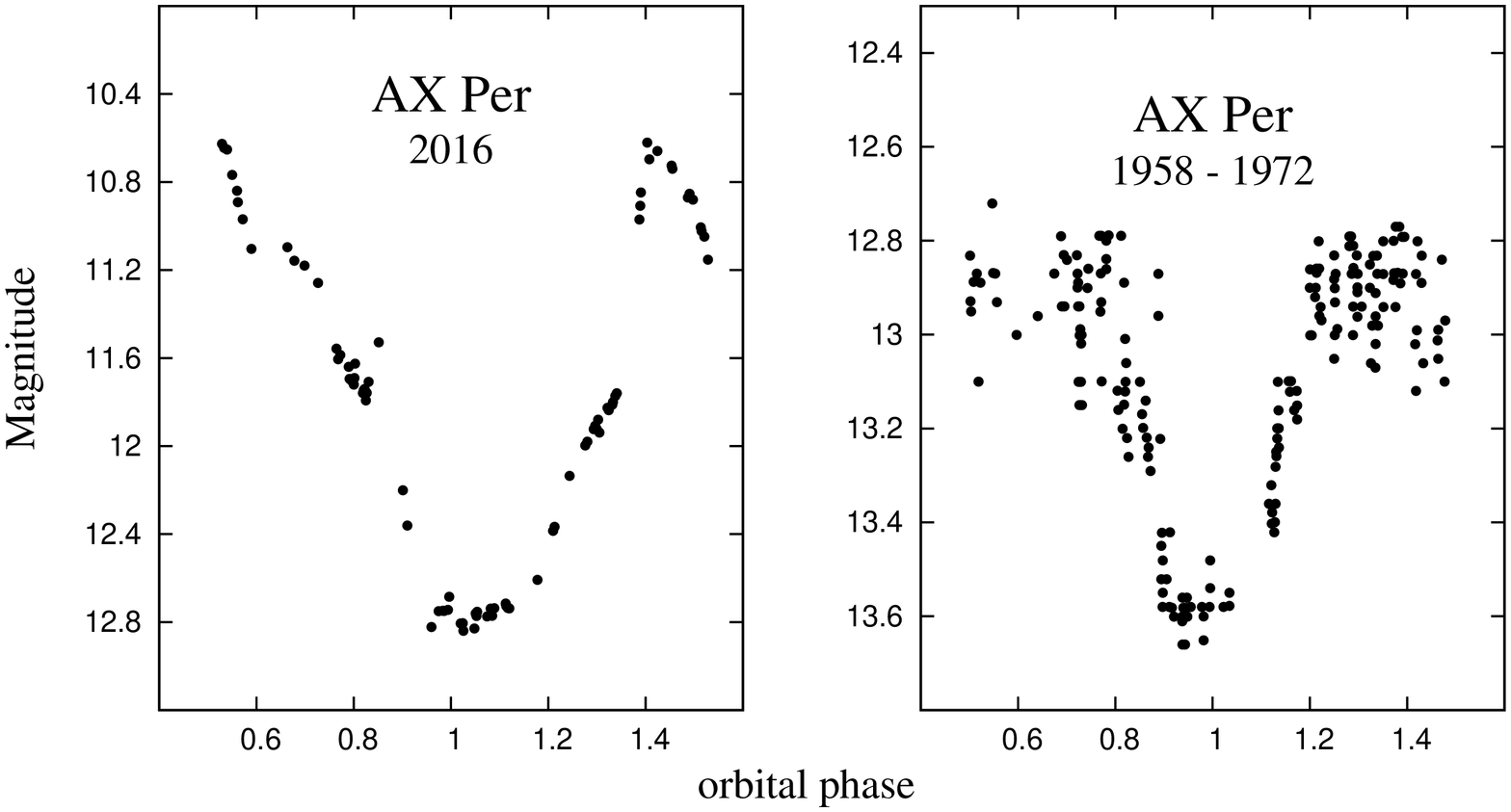}}
\caption{
A comparison between the shape of the 2016 eclipse during 
the current active phase and eclipses during the 1958-1972 
quiescent period \citep[data are from][]{1977PZP.....3...71M}. 
}
\label{axper_ecl}
\end{figure}
%
AX~Per is an eclipsing binary with the orbital period of 
$\sim 680$ days 
\citep[e.g.][]{1991IBVS.3603....1S,1992AJ....103..579M}. 
The last quiescent phase that lasted from 1995 was 
interrupted by the $\Delta U\sim0.5$\,mag flare in July 2007 
followed with rapid $\sim1$ and $\sim0.7$\,mag brightenings 
in $B$, during March 2009 and November 2010, respectively 
(Fig.~\ref{axper}), supporting a suggestion of a forthcoming 
major outburst 
\citep[][]{2009CBET.1757....1M,2010CBET.2555....1M}. 

Our new photometric observations cover nearly four orbital 
cycles during a high level of the star activity. A wave-like 
profile in all colours with the amplitude of $\sim2$\,mag are 
typical for each cycle (Figs.~\ref{axper} and \ref{axper_sep}). 
It is of interest to note that such characteristics can 
be recognized also in LCs of other active symbiotic binaries. 
Figure~\ref{outbursts} shows a comparison for AX~Per, BF~Cyg, 
Z~And and AR~Pav. 
The light minimum is always observed around the inferior 
conjunction of the giant, i.e. at the orbital phase 
$\varphi\sim0$, and the maximum light around $\varphi\sim0.5$, 
when the active object is in front. 

During the recent active phase, the minima -- eclipses 
differ significantly from those observed during previous 
active phases \citep[see Fig.~2 of][]{2011A&A...536A..27S}. 
For example, the eclipse measured in March 2015 has a narrow, 
symmetric V-shaped profile in $B$ and $V$ with the minimum at 
$JD_{\rm Min}\sim 2\,457\,090.8$, while in $U$ the LC is 
rather flat. 
The following `eclipse' from December 2016 
($JD_{\rm Min}\sim 2\,457\,743$) is much broader and asymmetric 
in $B$, $V$ and $R_{\rm C}$ and even flatter in $U$, 
lasting from $\varphi\sim0.9$ to $\sim1.2$ 
(Fig.~\ref{axper_sep}). We note here that a very similar eclipse 
profile was observed between 1958 and 1972 (Fig.~\ref{axper_ecl}), 
when AX~Per was in quiescent phase 
\citep[][]{1977PZP.....3...71M,2001A&A...367..199S}. 
According to $V$, $R_{\rm C}$ and $I_{\rm C}$ LCs, the last 
observed minimum started in August 2018 ($\varphi\sim0.86$) 
showing a broad descending branch. Although our observations 
end before the predicted eclipse time, it seems that the light 
minimum precedes this position (Fig.~\ref{axper_sep}). 
In general, the LC profile changes from cycle to cycle and 
is dependent on the colour. 
Such the behaviour reflects complex and variable sources of 
radiation around the WD as well as within 
the circumbinary environment. 
%
\subsection{FG~Ser}
\label{s:fgser}
%
%
\begin{figure}[p!t]
\begin{center}
\resizebox{\hsize}{!}{\includegraphics[angle=-90]{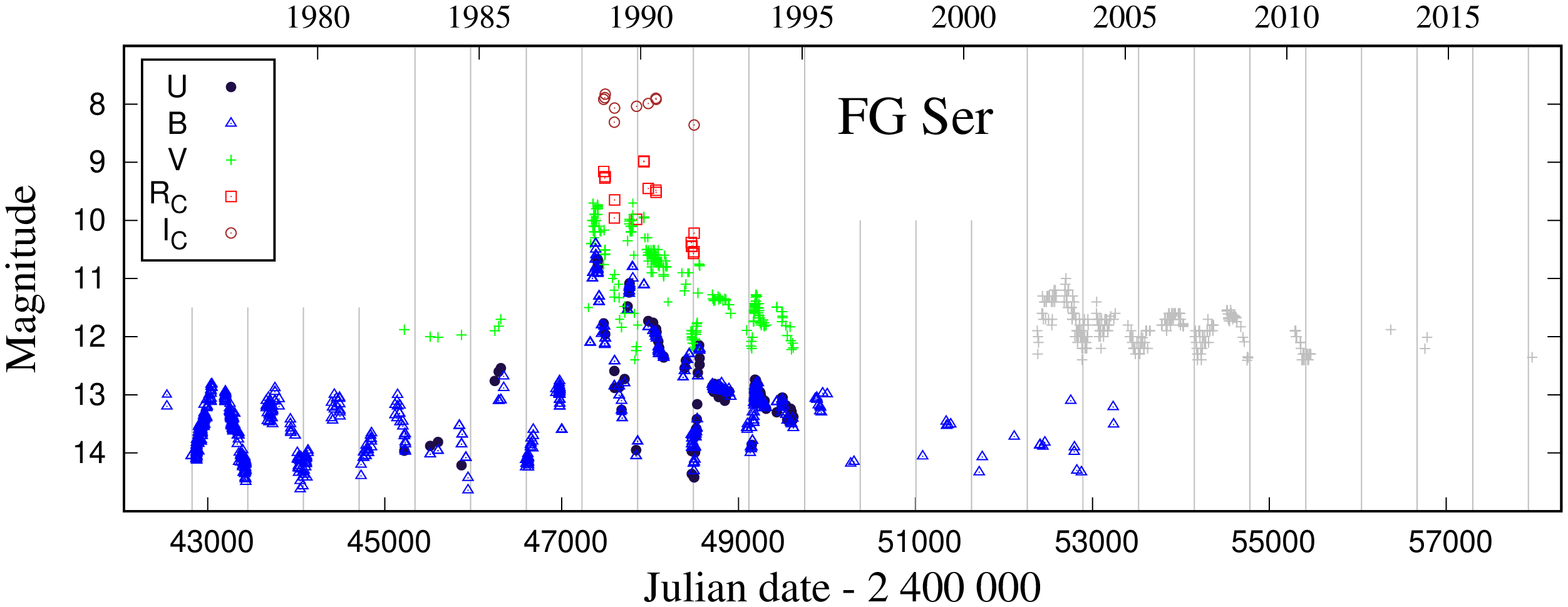}}\vspace*{4mm}
\resizebox{\hsize}{!}{\includegraphics[angle=-90]{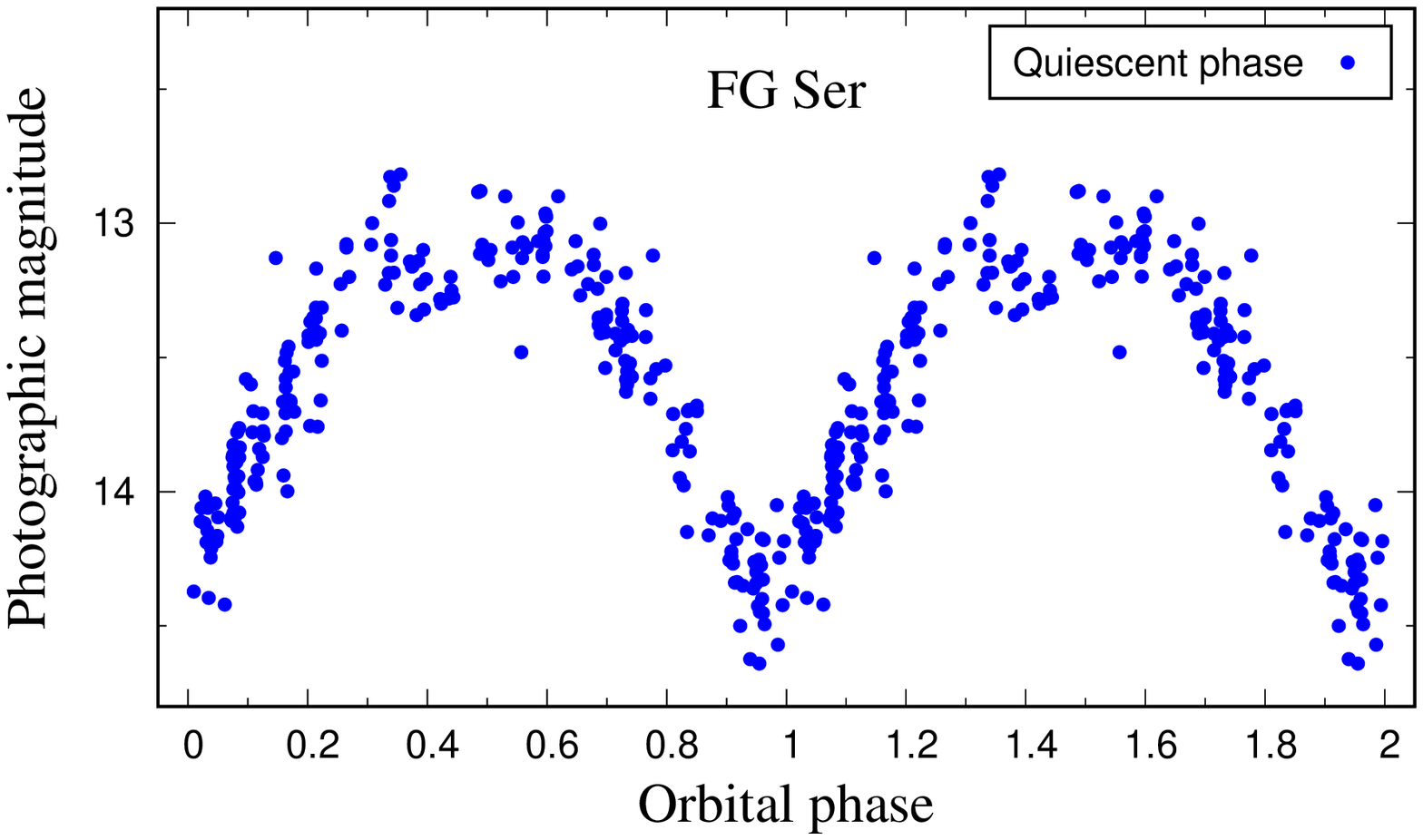}\hspace*{4mm}
                      \includegraphics[angle=-90]{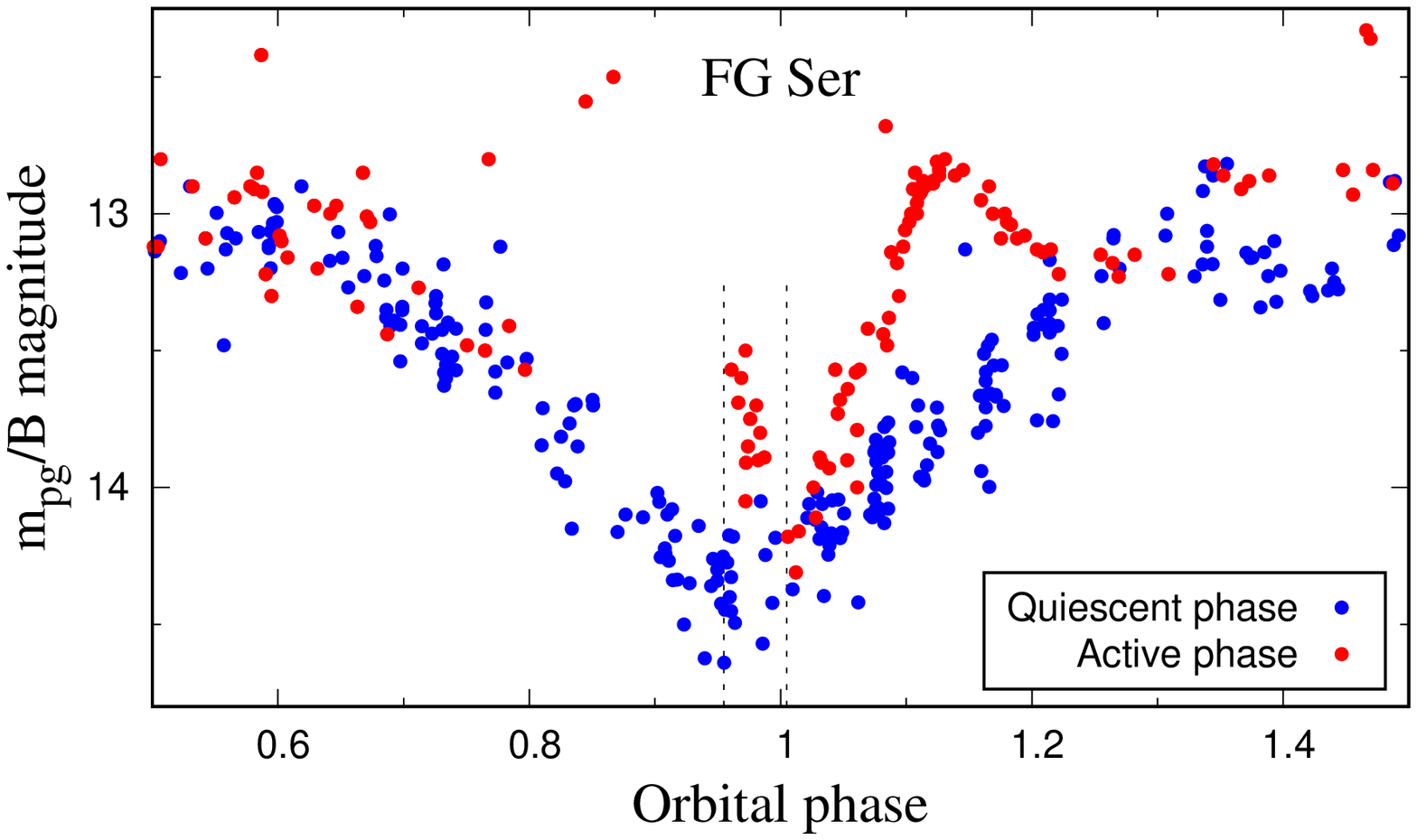}}
\end{center}
\caption{
Top: As in Fig.~\ref{egand}, but for FG~Ser. The vertical lines 
represent times of minima according to Eq.~(\ref{fgeph}). 
The figure displays our photographic magnitudes from archives, data 
published by \cite{1993A&AT....3..295K} (shifted by -0.2\,mag) 
and photographic, photoelectric and visual data of 
\cite{1992AJ....104..262M,1995AJ....109.1740M} (other colours). 
Compared are visual estimates from \textsl{AAVSO} (grey). 
Bottom: phase diagram of $m_{\rm pg}$ magnitudes during quiescent 
phase (left; $JD < 2\,447\,000$) and its comparison with $B$ 
measurements from the 1988-94 active phase (right). 
}
\label{fgser}
\end{figure}
Variability of the symbiotic star FG~Ser was discovered by 
\cite{1968AN....290..277H}. The star underwent an outburst 
in 1988 when its brightness increased by $\Delta B\sim3.5$\,mag 
and peaked at $B\sim 10$ on July 4, 1988 
\citep[][]{1988IAUC.4622....2M}. The next monitoring of the 
star's brightness showed multiple maxima and minima - eclipses 
\citep[see][and Fig.~\ref{fgser} here]{1995AJ....109.1740M}. 
Using the timing of three eclipses observed during 1989 -- 1993 
the authors determined the eclipse ephemeris as
$JD_{\rm Ecl.} = 2\,448\,492(\pm 4) + (658\pm 4)\times E$. 
%

During the quiescent phase, prior to the 1988 outburst, 
\cite{1993A&AT....3..295K} determined the ephemeris for 
regular brightness variations (B = 13--14.5) to 
$JD_{\rm Min} = 2\,446\,590.9 + 630\times E$ using 
the photographic plates from the Sternberg Astronomical 
Institute archive (SAI; 1949-1987). He also determined the 
ephemeris of the secondary minimum, which is observed at 
the phase $\varphi \sim0.45$ with the depth of $\sim$0.5\,mag. 
Using the same archive of photo-plates, but from 1901, the 
data of \cite{1992AJ....104..262M,1995AJ....109.1740M} and 
new photometric observations, \cite{2014aspl.conf..115S} 
rectified the minima ephemeris to 
\begin{equation}
  JD_{\rm Min} = 2\,443\,452(\pm 7) + (629.4\pm 1.0)\times E
\label{fgeph}
\end{equation}

For the purpose of this paper, we remeasured magnitudes of FG~Ser 
on photographic plates of the SAI archive (1955--1995; 256 plates)
and the Asiago archive (1968--1994; 41 plates). Our data are 
plotted together with those from literature in Fig.~\ref{fgser}. 
During the quiescent phase ($JD < 2\,447\,000$), their phase diagram 
with the ephemeris (\ref{fgeph}) shows typical wave-like 
orbitally-related variation with 
$\Delta m_{\rm pg} \sim 1.5$\,mag, broad minimum around the 
inferior conjunction of the giant and a shallow $\sim$0.3\,mag 
deep secondary minimum at $\varphi\sim 0.4$. During the active 
phase (1988-94), the minimum (eclipse) became narrower and shifted 
its position after that from the quiescent phase by 
$\sim 0.05\times P_{\rm orb}$ (bottom right panel of the figure), 
which prolongs the minima separation during the transition from 
quiescence to activity and during the decline from the outburst. 
This effect of apparent changes of orbital periods in symbiotic 
binaries was revealed by \cite{1998A&A...338..599S}. Here, we 
indicated this effect also for AE~Ara (see Sect.~\ref{ss:aeara}). 
%
%
%
\begin{figure}[p!t]
\begin{center}
\resizebox{\hsize}{!}{\includegraphics[angle=-90]{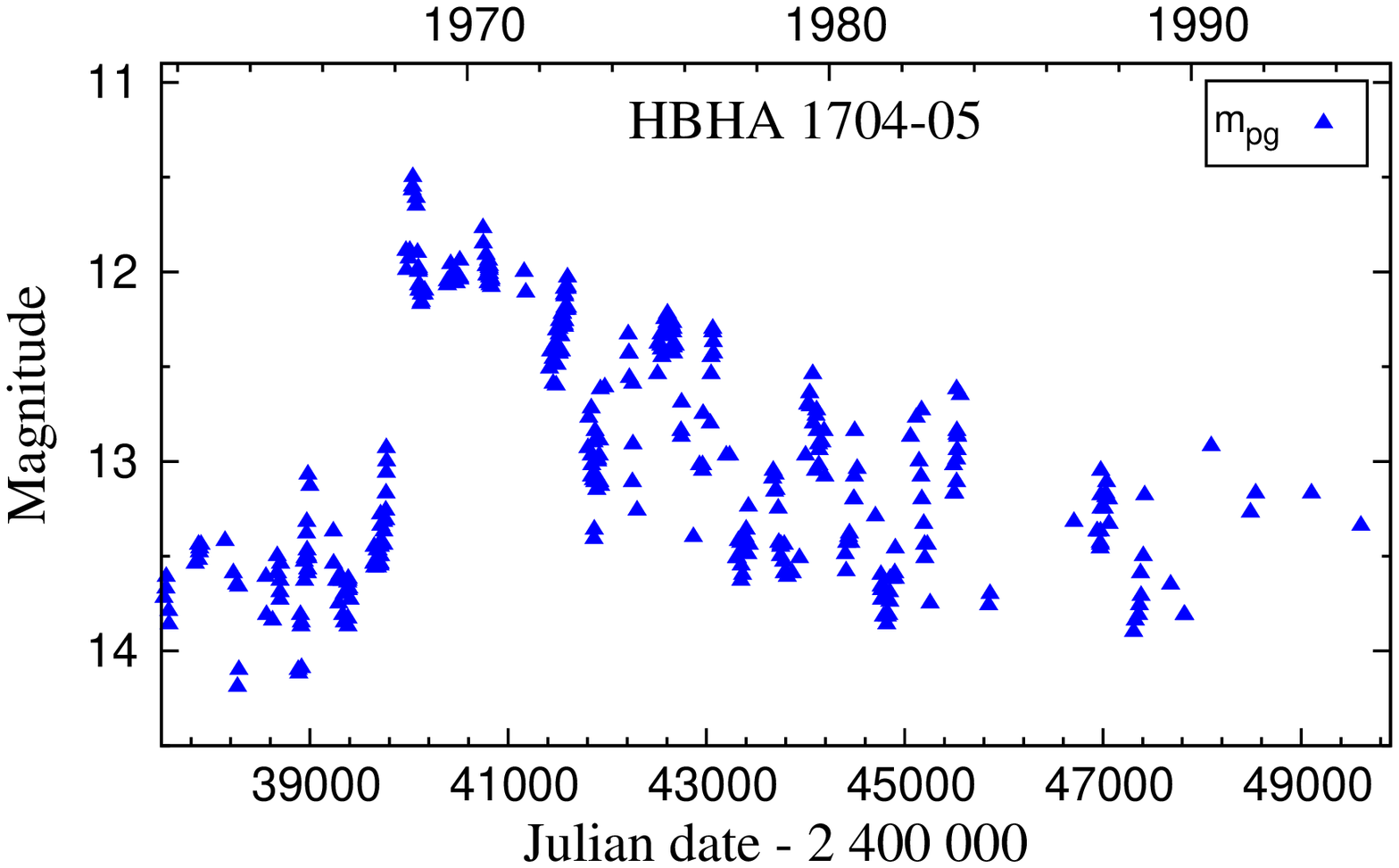}\hspace*{2mm}
                      \includegraphics[angle=-90]{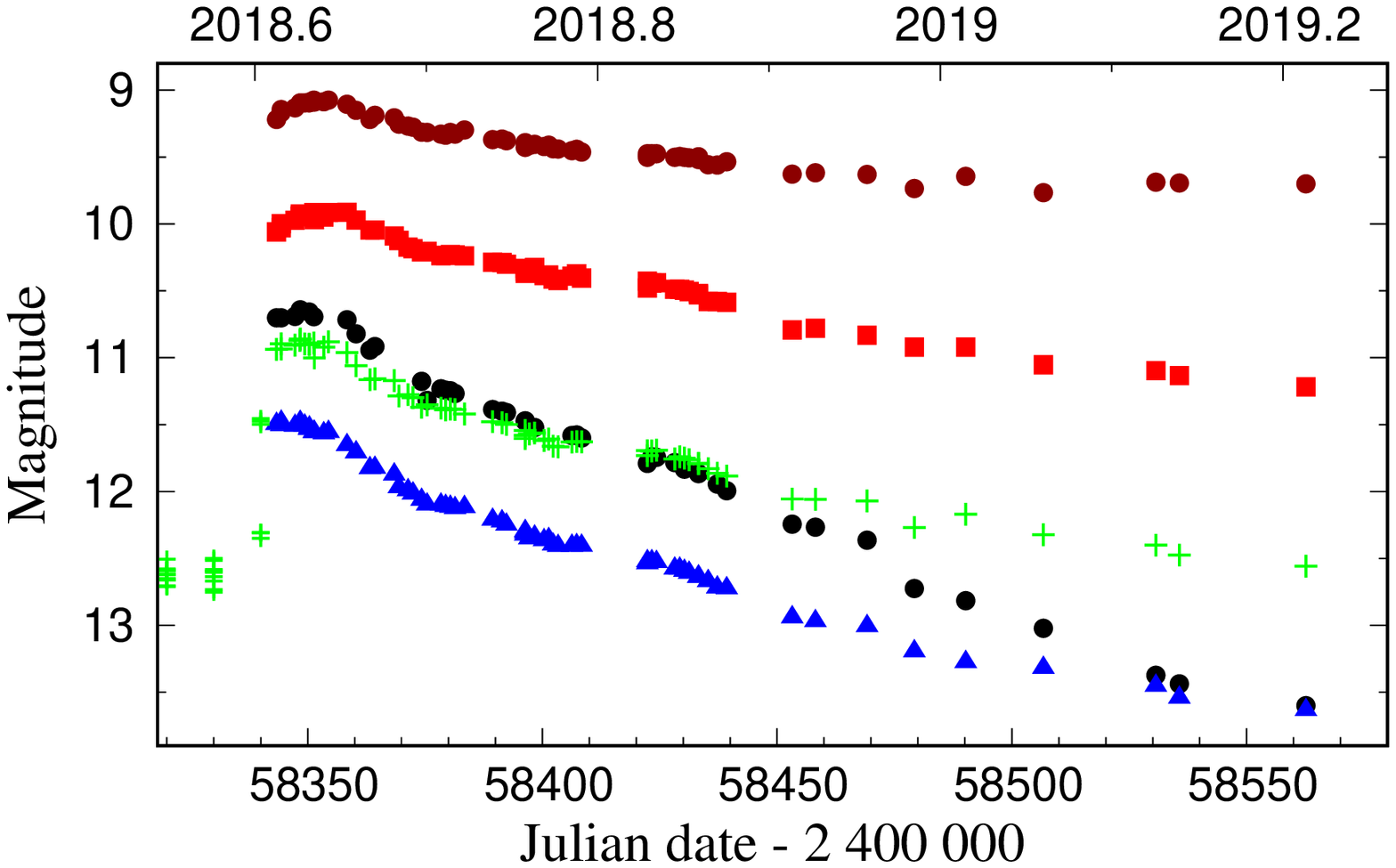}}
\end{center}
\caption{
Left: Historical LC of HBHA~1704-05 obtained from the Moscow's 
photographic plates archive. 
Right: Our $UBVR_{\rm C}I_{\rm C}$ LCs measured from the maximum 
of the outburst, on August 12, 2018. Compared are the \textsl{ASASSN} 
$V$ magnitudes prior to the outburst. Denotation of points as in 
Fig.~\ref{egand}. 
}
\label{fig:hbha}
\end{figure}

\subsection{HBHA~1704-05}
\label{s:hbha}
The star HBHA 1704-05 was originally classified as an emission-line 
star \citep[][]{1999A&AS..134..255K}. On the basis of The All Sky 
Automated Survey for SuperNovae (ASAS-SN), the object was catalogued 
in the VSX as a semi-regular variable with a periodicity around 418 
days. On August 11, 2018, \citet{2018ATel11937....1M} reported that 
HBHA~1704-05 is coincident with the star TCP J19544251+1722281 
(ASASSN-V J195442.95+172212.6), which brightened from $V = 12.0$ 
on July 31.945, 2018 to $V = 10.7$ on August 8.938, 2018, showing 
characteristics of a symbiotic star in outburst.\footnote{
http://www.cbat.eps.harvard.edu/unconf/followups/J19544251+1722281.html}
In the spectrum, they found features of an M-type giant, a strong 
nebular continuum with superposed emission lines of highly ionized 
elements (e.g., He\,{\small II}\,4686\,\AA\ and 
[Ne\,{\small V}]\,3426\,\AA), confirming the nature of HBHA~1704-05 
as a symbiotic star in outburst. Photometric and spectroscopic 
monitoring of HBHA~1704-05 during the first month of its outburst 
was presented by \cite{2019CoSkaS}. 

We started to monitor HBHA 1704-05 on August 12, 2018, just after 
the discovery of its outburst. To obtain absolute values of its 
$UBVR_{\rm C}I_{\rm C}$ magnitudes we used standard stars '120' 
and '123' as indicated on the \textsl{AAVSO} chart and the check star 
C = TYC~1620-2479 ($\alpha_{2000} = 19^{\rm h}54^{\rm m}55^{\rm s},~
\delta_{2000} = +17^{\circ}23^{\prime}36^{\prime\prime}$). 
Using the standard stars $\alpha$, "b", "c" in the field of 
PU~Vul and "a", "b", "c" in the field of HM~Sge 
\citep[see charts of][]{2006A&A...458..339H}, we determined 
the $UBVR_{\rm C}I_{\rm C}$ magnitudes of the standard stars 
in the field of HBHA~1704-05. Resulting magnitudes represent 
mean values with $rms$ errors obtained during 5 nights 
(see Table~\ref{tab:comp}). 

Our photometry of HBHA 1704-05 is depicted in Fig.~\ref{fig:hbha}. 
The right panel of the figure shows 
our multicolour CCD photometry during the current outburst. 
The LCs show a gradual decline, steepness of which depends 
on the colour. From the August's 2018 maximum to our last 
measurements during February 2019, the star's brightness decreased 
by $\sim$2.6, $\sim$2.0, $\sim$1.5, $\sim$1.1 and $\sim$0.6\,mag 
in the $U$, $B$, $V$, $R_{\rm C}$ and $I_{\rm C}$ filter, 
respectively. 
We also measured photographic magnitudes on the plates collected 
in the Moscow's photographic plates archive of the Sternberg 
Astronomical Institute of Moscow State University. 
These measurements revealed another 2-mag outburst lasting 
from 1968 to $\sim$1975 (left panel of the figure). 
A detailed analysis of LCs together with spectroscopic 
observations of HBHA~1704-05 is planned for another 
publication. 
\section{Conclusions}
\label{s:concl}

In this paper, we present new multicolour long-term photometry 
of selected symbiotic stars. Main results of our monitoring 
program can be summarized as follows. 

\textbf{EG\,And} persists in a quiescent phase. Our observations 
confirmed a double wave variation along the orbit in all 
$UBVR_{\rm C}I_{\rm C}$ LCs. We also detected several light 
variations, with the most significant period around 28 days 
in the $B$ LC. 

\textbf{Z\,And} continues its active phase that started during 
September 2000. 
The amplitude of the brightness peaks, which appeared around 
$\varphi = 0.5$ was gradually decreasing from $U\sim8.3$ in 
November 2011, $\sim9.1$ between February and May 2014 to 
$\sim9.6$\,mag in June 2016. The recent outburst began 
in October 2017, peaking at $U\sim8.5$\,mag in April 2018. 
We also indicated a periodic light variation with 
a $\sim58$-day period, visible in our $UBV$ data. 

%
\textbf{AE\,Ara} came to quiescent phase from around 2016. 
During its recent active phase, from 2006 to 2016, we 
indicated a period of 850--860 days, which is distinctive 
larger than its orbital period of $812\pm 2$ days from 
light variations during preceding quiescence or $803\pm 9$ 
days from radial velocities. Such an apparent change 
in the orbital period is caused by the change of the 
ionization structure of the binary during quiescent and 
active phase \citep[see][]{1998A&A...338..599S}. 

\textbf{BF\,Cyg} remains at a high level of activity since 
its major outburst in 2006. The overall star's brightness 
was gradually declining until the eclipse position in 2014, 
when the trend turned into the opposite way, with the 
increasing delay of the minima position. During 2014, 2016 
and 2018 the light minimum delayed by $\sim$14, $\sim$17 and 
$\sim$20 days after the inferior conjunction of the giant. 
The recent 2018-minimum ($JD_{\rm Min}\sim2\,458\,230$) was 
followed by a $\sim0.3$\,mag brightening peaked at 
$U\sim9.7$\,mag on $JD_{\rm Max}\sim2\,458\,360$. 

\textbf{CH\,Cyg} continues its active phase showing extremely 
complex photometric variability. 
Its brightness was gradually increasing to 2015 when reached 
a maximum of $U\sim7$\,mag.  
During the following descending part of the $UB(V)$ LCs 
CH\,Cyg underwent three bursts with a comparable amplitude 
of $U\sim1.7$\,mag. During 2016, the light minima were shifted 
with increasing wavelength. Then CH\,Cyg reached its maximum 
at $U\sim 6.4$\,mag during the first half of 2018, being 
followed with a sharp minimim in May 2018 and continuing 
a $\sim2$\,mag decrease in $U$ by the end of 2018 
(see Fig.~\ref{ch_close}). 

\textbf{CI\,Cyg} experienced its last activity in the form 
of a double-peaked burst at the end of 2012. Then entered 
a quiescent phase showing pronounced orbitally-related light 
variation in the LCs. The recent broad minima in 2015 and 
2017 differ in their position and profile for different colours. 
Secondary minima around $\varphi\sim$0.6 are present in 
$BVR_{\rm C}I_{\rm C}$ LCs. A $\sim73$ days light variation 
can be recognized in the $V$ LC 
(see Figs.~\ref{cicyg_ecl} and \ref{cicyg}). 

\textbf{V1016\,Cyg} 
continued a slow brightness decline in 
$UBVR_{\rm C}$ LCs by a few times 0.1\,mag during our observing 
period (from March 2016), whereas the $I_{\rm C}$ LC reflects 
pulsations of the Mira-type giant with 
$\Delta I_{\rm C}\sim$0.7\,mag. 
A sudden drop in the $U$ magnitude by $\sim0.17$\,mag was measured 
around 2015.4 (see Fig.~\ref{v1016s}), confirming the previous 
finding by \cite{2016BaltA..25...35A,2016yCat..90410665A}. 
A decrease in the $U-B$ and $B-V$ colours (Fig.~\ref{v1016_CI}) 
can be in part caused by the overall star's brightness decline 
and a flux decrease of strong emission lines contributing to 
$V$ and $B$ passbands. 

\textbf{V1329\,Cyg} persists in quiescent phase showing well 
pronounced orbitally-related light variation with a decreasing 
amplitude at longer wavelengths. Their minima precede 
the inferior conjunction of the giant by 
$\sim 0.1\times P_{\rm orb}$ (Fig.~\ref{fig:v1329}). 

\textbf{AG\,Dra} continued its quasi-quiescent phase, during 
which the wave-like variability is occasionally interrupted by 
short-lasting ($\lesssim$3 months) bursts with 
$\Delta U < 2$\,mag. Bright bursts usually show a multi-peaked 
structure (see Figs.~\ref{agdra} and \ref{agdra_hot}). 
We noticed a rather constant $U-B$ colour during the last 2006-08 
major outburst, while the $B-V$ index does not show such 
feature. Similar plateau in the $U-B$ LC was probably 
present also during the 1980-82 and 1994-95 major outbursts 
(see Fig.~\ref{agdra_plat}). 

\textbf{RS\,Oph} persists at the stage between its recurrent 
outbursts. The $BVR_{\rm C}I_{\rm C}$ LCs from the last 2006 
eruption to the present suggest a smooth variation with the 
amplitude $\Delta V\sim0.3$\,mag and a period of $\sim$9.4 
years (Fig.~\ref{rsoph}). Rapid variability is detected in 
all filters. High-cadence measurements in $B$ demonstrates 
a stochastic variability with $\Delta B\sim 0.5$\,mag on 
the time-scale of hours and $\Delta B\sim 0.05-0.1$\,mag on 
the time-scale of minutes (Fig.~\ref{rsoph_flick}). 

\textbf{AR\,Pav} reduced its out-of-eclipse activity from 
around 2004, showing only two $\sim0.5$\,mag brightenings 
in 2013 and 2014 (Fig.~\ref{arpav}). We determined positions 
of new minima from the epoch 71 to 78. All the available minima 
(1896.3 -- 2018.8, Table\ref{tpav}) determine the linear 
ephemeris, 
$JD_{\rm Min} = 2\,411\,264.4(\pm 1.7) + (604.47\pm 0.03)\times E$. 
The corresponding $O-C$ diagram indicates real systematic 
variation in the minima timing (Fig.~\ref{fig:ar_eph}). 

\textbf{AG\,Peg} 
underwent its first Z~And-type outburst in June 2015, after 
165 years from its nova-like outburst in 1850 (Fig.~\ref{agpeg}). 
This change indicates the beginning of a new era in the AG~Peg 
evolution. 
After about 1 year AG~Peg came to a quiescent phase. 

\textbf{AX\,Per} 
persists in a low-level active stage from about 2007. 
The LC profile varies from cycle to cycle and depends 
on the colour. For example, in 2016, the maximum in $U$ 
preceded those in other filters by $\sim48$ days, while 
in 2017, maxima in $U$ and $B$ occurred at the same time, 
around $\varphi = 0.4$. The $R_{\rm C}$ LC shows 
secondary minima around $\varphi = 0.45$. 

\textbf{FG\,Ser} was in quiescent phase from 1901 to 1987 
as documented by re-measuring its archival photographic 
measurements. We determined a new ephemeris for its minima 
times, 
$JD_{\rm Min}=2\,443\,452(\pm 7)+(629.4\pm 1.0)\times E$. 
During the active phase (1988-94), the minimum (eclipse) 
occurred by $\sim 0.05\times P_{\rm orb}$ after the minima 
positions from quiescence. 

\textbf{HBHA~1704-05} is the newly (August 9, 2018) discovered 
symbiotic star in outburst. Our $UBVR_{\rm C}I_{\rm C}$ photometry 
shows that the current outburst is of Z~And-type. By inspection 
of the Moscow's archive of photographic plates, we revealed another 
2-mag-outburst of HBHA~1704-05 in 1968.

\acknowledgements
We thank the anonymous referee for constructive comments. 
Evgeni Kolotilov, Kirill Sokolovsky and Alexandra Zubareva 
are thanked for their assistance in acquisition of some data 
from the Moscow's photographic plates archive. Alisa Shchurova 
is thanked for her assistance in carrying out some CCD 
observations in the G2 pavilion within her PhD study. 
This work was supported by the Slovak Academy of Sciences grant 
VEGA No. 2/0008/17, by the Slovak Research and Development Agency 
under the contract No. APVV-15-0458 and by the Bulgarian Scientific 
Research Fund of the Ministry of Education and Science under 
the grants DN 08-1/2016 and DN 18-13/2017. 
Sergey Shugarov acknowledges the support from the Program of 
development of M.V. Lomonosov Moscow State University (Leading 
Scientific School 'Physics of stars, relativistic objects and 
galaxies'). 
We also acknowledge with thanks the variable star observations 
from the AAVSO International Database contributed by observers 
worldwide and used in this research.


\bibliography{ss14}


\end{document}